 \documentclass[final,3p,times]{elsarticle}

\usepackage{amssymb}

\usepackage{lineno}

\journal{Journal Name}

\newtheorem{thm}{Theorem}
\newtheorem{lem}{Lemma}
\newdefinition{rmk}{Remark}

\newtheorem{prop}{Proposition}
\newproof{pf}{Proof}
\newproof{pot}{Proof of Theorem \ref{thm2}}

\begin{document}

\begin{frontmatter}

\title{ Strongly Consistent of Kullback-Leibler Divergence Estimator and Tests for Model Selection Based on a Bias Reduced Kernel Density Estimator}

\author[mymainaddress]{Papa NGOM }
\cortext[mycorrespondingauthor]{Corresponding author: papa.ngom@ucad.edu.sn (Papa NGOM) }
\author[mysecondaryaddress]{Freedath DJIBRIL MOUSSA }
\author[mythirdlyaddress]{Jean de Dieu NKURUNZIZA}
\address[mymainaddress]{LMA,Universit\'{e} Cheikh Anta Diop, Dakar, Senegal}
\address[mysecondaryaddress]{Facult\'e des Sciences et Techniques, Abomey-Calavi, Benin}
\address[mythirdlyaddress]{Institut de Math\'ematiques et de Sciences Physiques, Porto Novo, Benin}

\begin{abstract}
In this paper, we study the strong consistency of a bias reduced kernel density estimator
and derive a strongly consistent Kullback-Leibler divergence (KLD) estimator. As application, we formulate a goodness-of-fit test and 
an asymptotically standard normal test for model selection. The Monte Carlo simulation 
show the effectiveness of the proposed estimation methods and statistical tests.
\end{abstract}

\end{frontmatter}


\section{Introduction}

Let $ X_{1},...,X_{n} $ be iid random variables and assume that the common distribution
function of these variables has an unknown density $ f $. One can estimate $ f $ using the parametric approach assuming 
 that the data are drawn from a known parametric family of distributions. The density $ f $ can then be estimated by
 finding estimates of the parameters from the data and substituting these estimates into the formula of the density.
One can also use non-parametric approach for the density estimation. A well known non-parametric
estimator of the pdf (probability density function) is the histogram \cite{30}. It has the advantage of simplicity but it also has
some disadvantages, such as: lack of continuity and the choice of the location of intervals and the bandwidth
have an effect on the histogram result.
To circumvent such difficulties, Rosenblatt and Parzen \cite{30,31} proposed a more general 
non-parametric estimator which is the widely used kernel density estimator. 
The asymptotic properties of this estimator has been intensively investigated and many kernel-types estimators have been proposed. Dony and Einmahl \cite{SI} showed the uniform consistency of kernel density
estimator with general bandwidth sequences. Salim and Issam \cite{SBI} established the uniform in bandwidth consistency of kernel-type estimators
of Shannon Entropy. Einmah and Mason \cite{EMD} proved the uniform in bandwidth consistency of 
kernel-type function estimators. Ngom et al. \cite{PHD} proposed a strong uniformly consistent kernel-type estimator of divergence measures. 
Xie and Wu \cite{29} focused on improving convergence rate of kernel density estimator
by introducing a bias reduced kernel density estimator. The first main purpose of this paper is to prove the strong consistency of this bias reduced kernel density estimator. To choose a pratical optimal bandwidth of classical kernel density estimator, one way of determining a simple and 
attractive smoothing parameter is the cross-validation method introduced by Rudemo and Bowman \cite{TB}. 
Accordingly we shall propose a cross-validation bandwidth selection
for the bias reduced kernel density estimator.
Next, we adress the model selection problem. Considering a candidate model for some given data generated by an unknown probability distribution, the 
dissimilarity between those two probability distributions can be measured by the Kullback-Leibler divergence (KLD) introduced by  Kullback and 
 Leibler \cite{SR}. 
 Since the true density is unknown, various criteria and hypothesis testing were used for model selection purpose (
\cite{19,EMH, EEH, 4, KDA, KDA2, ASK, QV, QV1}).
In this paper, we shall derive a strongly consistent estimator of
 KLD between two distributions based on the bais reduced kernel density estimator. 
 The proposed KLD estimator is then used to construct statistics for hypothesis testing in model selection. \bigskip

The rest of the paper is organized as follows. We give a brief review of the bias reduced kernel density estimator in Section 2.  
Cross-validation bandwidth selection for the bias reduced kernel density estimator is obtained in Section 3. 
In Section 4, the strong consistency of the bias reduced kernel density estimator is proved and 
we establish a strongly consistent  Kullback-Leibler divergence estimator  in Section 5.
Applications for hypothesis testing in models selection are proposed in Section 6. 
The simulation study is presented in Section 7 and finally the conclusion appears in Section 8.

\section{ A  review of  the bias reduced kernel density estimator}

Let $ X_{1},...,X_{n} $ be iid random variables and assume that the unknown distribution
function of these variables has a Lebesgue density, which we shall denote
by $f$. Consider a probability density function $K$ defined on $\mathbb{R}$ (the kernel) and a positive parameter $h$, the bandwidth. 
Assuming that the random variable of density $K$ is centered with finite variance $\mu_{2}$, 
the kernel density estimator (Rosenblatt \cite{30} and Parzen \cite{31}) of $f$ is given by 
\begin{eqnarray}\label{E1}
\hat{f}_{n,h}(x)=\frac{1}{nh}\sum_{i=1}^{n}K\left(\frac{x-X_{i}}{h} \right).
\end{eqnarray}
 Devroye \cite{33} showed that the optimal bandwidth 
is $ h\sim O\left( n^{-\frac{1}{5}}\right) $ and then the optimal MSE is of the order $ n^{-\frac{4}{5}}$, under the conditions
 $h\rightarrow 0$ and $nh\rightarrow \infty $ as $n\rightarrow \infty $.\\

The optimal performance of kernel density estimator has been widely investigated. Farrell \cite{FR} obtained the best
asymptotic convergence rate of MSE for orthogonal kernel estimators.
 Abramson \cite{AI} successfully employed larger smoothing parameters in low density regions 
to reduce the bias. Samiuddin \cite{SE} reduced bias by introducing the idea of inadmissible kernels. 
On the other hand, El-Sayyad and Samiuddin in \cite{ESA} used some probabilistic arguments in proposing an estimator 
which achieves the goal of bias reduction. Ruppert and Cline \cite{RC} used the estimated c.d.f. in introducing a bias reduction 
method via smoothed kernel transformations. Mynbaev and Martius \cite{MM} used the Lipschitz condition in 
 order to work out a bias reduced kernel relative to 
the classical kernel estimator. Kim \cite{KK} reduced the bias and variance at the same time 
which in turn reduces the MSE in using the skewing method. Xie and Wu \cite{29} proposed a very intuitive and
feasible kernel density estimator which reduces the bias and MSE significantly compared to the ordinary kernel density estimator.
It is defined by
\begin{eqnarray}\label{E2}
\hat{f}_{n,h}^{b}(x) =\hat{f}_{n,h}(x)-\frac{h^{2}}{2}\mu_{2}\hat{f}_{n,h}^{ ''}(x).
\end{eqnarray}

Assuming that $f$ is differentiable of order four in a neighbourhood of $x$, Xie and Wu \cite{29}
came up with a convergence rate of order $n^{-\frac{6}{7}}$ for this estimator. We prove this resultats under the appropriate regularity 
 conditions on the kernel $K$.
\begin{prop} 
\label{thm1}
Suppose that $f$ is differentiable of order four in a neighbourhood of $x$. Let $ K $ be the density of a centered random variable with finite second and third order moment denoted by $\mu_{2}$ 
and $\mu_{3}$ respectively, satisfying the following assumptions:
\begin{itemize}
\item $A_1$ : $K(x)=K_{1}(x)1_{A}(x),~A\subseteq \mathbb{R}$ such that 
$ \lim_{\stackrel{x\rightarrow \inf A }{ x> \inf A } }K^{(i)}(x)=\lim_{\stackrel{x\rightarrow \sup A }{ x< \sup A } }K^{(i)}(x)=0,~\forall i=0,1 $;
\item $A_2$ : $\int K^2(u)du<\infty$; $\int (K^{''}(u))^2du<\infty$; $\int u(K^{''}(u))^2du=0.$
\end{itemize} Then, we have 
\begin{eqnarray}
Bias(\hat{f}_{n,h}^{b}(x))=-\frac{h^{3}}{6}\mu_{3}f^{'''}(x)+O(h^{4})
\end{eqnarray}
and 
\begin{eqnarray}\label{V}
var(\hat{f}_{n,h}^{b}(x)) \leq \frac{1}{2nh}  \mu_{2}^{2} f(x) \int ( K^{''}(u))^{2}du+ O((n)^{-1}).
\end{eqnarray}
Consequently the optimal MSE (Mean Squared Error) is of the order $  n^{-\frac{6}{7}}  $.
\end{prop}
 If in addition, $K$ is a symetric kernel, $\mu_3=0$; hence the optimal MSE for the bias reduced estimator is of order $n^{-8/9}$.
 \begin{pf}
 $ Bias(\hat{f}_{n,h}^{b}(x) )= \mathbb{E} \hat{f}_{n,h}^{b}(x)-f(x),\forall x\in \mathbb{R} $
 where $\hat{f}_{n,h}^{b}(x) = \hat{f}_{n,h}(x) -\frac{h^{2}}{2}\mu_{2}\hat{f}_{n,h}^{ ''}(x)$
and $\hat{f}_{n,h}$ is a kernel density estimator, then
 \begin{eqnarray}\label{1}
  Bias(\hat{f}_{n,h}^{b}(x))&=&   \mathbb{E} \hat{f}_{n,h}(x)-\frac{h^{2}}{2} \mu_{2} \mathbb{E} \hat{f}_{n,h}^{ ''}(x) -f(x).
 \end{eqnarray}
 We have
 \begin{eqnarray*}
  \mathbb{E} \hat{f}_{n,h}(x)&=& \mathbb{E} \left[ \frac{1}{nh}\sum_{i=1}^{n}K\left( \frac{x-X_{i}}{h}\right)\right]\\
                             &=& \frac{1}{h}\int K\left( \frac{x-y}{h} \right) f(y)dy\\
                             &=& \int K\left(u \right)f(x-uh)du.
 \end{eqnarray*}
A Taylor expansion of  $ f(x-uh) $ yields
\begin{eqnarray*}
f(x-uh)=f(x)-uhf^{'}(x) + \frac{1}{2}(uh)^{2}f^{''}(x)-\frac{(uh)^{3}}{6}f^{'''}(x)+o(h^{3}).
\end{eqnarray*}
Thus 
\begin{eqnarray*}
 \mathbb{E}  \hat{f}_{n,h}(x) &=&  f(x)\int K(u)du-hf^{'}(x)\int u K(u)du  +\frac{h^{2}}{2}\mu_{2} f^{''}(x) -\frac{h^{3}}{6}\mu_{3}f^{'''}(x)+o(h^{3})\\\nonumber
                              &=& f(x)+\frac{h^{2}}{2}\mu_{2}f^{''}(x)-\frac{h^{3}}{6}\mu_{3}f^{'''}(x)+o(h^{3}).
\end{eqnarray*}
 On the other hand,
\begin{eqnarray*}
 \mathbb{E} \hat{f}_{n,h}^{ ''}(x) &=&\mathbb{E} \left[ \frac{1}{nh^{3}}\sum_{i=1}^{n}K^{''}\left( \frac{x-X_{i}}{h}\right)\right]\\
                                &=& \frac{1}{h^{3}}\int K^{''}\left( \frac{x-y}{h}\right)f(y)dy\\
                                &=& \frac{1}{h^{2}} \int K^{''}(u)f(x-uh)du.
\end{eqnarray*}
By integrating by part twice, and using assumption $A_{1}$, we get
\begin{eqnarray*}
 \mathbb{E} \hat{f}_{n,h}^{ ''}(x) = \int K(u)f^{\ ''}(x-uh)du.
 \end{eqnarray*}                                 
Using a 2nd order Taylor expansion of $f^{\ ''}(x-uh)$ about $x$ we have
 \begin{eqnarray*}
 \mathbb{E} \hat{f}_{n,h}^{ ''}(x) &=& \int K(u)\left[ f^{''}(x)-uhf^{'''}(x)+\frac{(uh)^{2}}{2}f^{''''}(x)+ o(h^{2})\right] du \\
                                  &=& f^{''}(x)\int K(u)du -hf^{'''}(x)\int u K(u)du+\frac{h^{2}}{2}\mu_{2}f^{''''}(x) + o(h^{2})\\
                                  &=&f^{''}(x)+\frac{h^{2}}{2}\mu_{2}f^{''''}(x) + o(h^{2}).
 \end{eqnarray*}
Thus
\begin{eqnarray*}
Bias(\hat{f}_{n,h}^{b}(x) )&=& \mathbb{E} \hat{f}_{n,h}^{b}(x)-f(x) \\
             &=& \mathbb{E}(\hat{f}_{n,h}(x))-\frac{h^{2}}{2}\mu_{2} \mathbb{E} \left( f^{''}_{n}(x) \right)-f(x) \\
             &=& \frac{h^{2}}{2}\mu_{2}f^{''}(x) -\frac{h^{3}}{6}\mu_{3}f^{'''}(x)+o(h^{3})
             -\frac{h^{2}}{2}\mu_{2}\left[f^{''}(x) +\frac{h^{2}}{2}\mu_{2}f^{''''}(x) + o(h^{2})\right]\\
             &=& -\frac{h^{3}}{6}\mu_{3}f^{'''}(x)-\frac{h^{4}}{4}\mu_{2}^{2}f^{''''}(x)+o(h^{3}).
 \end{eqnarray*}            
Hence
\begin{eqnarray*}
Bias(\hat{f}_{n,h}^{b}(x)) = -\frac{h^{3}}{6}\mu_{3}f^{'''}(x)+O(h^{4}).                
\end{eqnarray*}
Consider now 
\begin{eqnarray}\label{1}
var(\hat{f}_{n,h}^{b}(x))= var\left( \hat{f}_{n,h}(x)-\frac{h^{2}}{2}\mu_{2}\hat{f}_{n,h}^{''}(x) \right).
\end{eqnarray}
We have
\begin{eqnarray}\label{varr}
var(\hat{f}_{n,h}^{b}(x)) & \leq & 2 var\left( \hat{f}_{n,h}(x) \right) + 2 var\left( \frac{h^{2}}{2}\mu_{2}\hat{f}_{n,h}^{ ''}(x) \right)\\ \nonumber
                     &\leq & 2 var\left(  \hat{f}_{n,h}(x) \right) +\frac{h^{4}}{2}\mu_{2}^{2} var\left( \hat{f}_{n,h}^{ ''}(x)\right).
\end{eqnarray}
Notice that the variance of $ \hat{f}_{n,h}(x) $ is  given by 
\begin{eqnarray}\label{varR}
var (  \hat{f}_{n,h}(x)) = \frac{1}{nh}f(x)\int K(u)^{2}du+ o\left( (nh)^{-1} \right).
\end{eqnarray}
 Set $I=var \left(\hat{f}_{n,h}^{ ''}(x)\right) $, we have  
\begin{eqnarray*}
I &= & var\left(  \frac{1}{nh^{3}}\sum_{i=1}^{n}K^{''}\left( \frac{x-X_{i}}{h}\right)\right)\\
  &= &\frac{1}{nh^{6}}\left[ \mathbb{E}\left(\left( K^{''}\left( \frac{x-X_{1}}{h}\right)\right)^{2}\right)-\left( \mathbb{E}\left( K^{''}\left( \frac{x-X_{1}}{h}\right)\right) \right)^{2} \right] \\
  &= &\frac{1}{nh^{6}}\int \left( K^{''}\left( \frac{x-y}{h}\right)\right)^{2}f(y)dy - \frac{1}{n}\left[ \frac{1}{h^{3}}\int K^{''}\left( \frac{x-y}{h}\right)f(y)dy\right]^{2} \\
  &=&\frac{1}{nh^{5}}\int (K^{''}(u))^{2}f(x-uh)du-\frac{1}{n}\left[ f^{''}(x)+ O(h^{2})\right]^{2}\\
  &=&\frac{1}{nh^{5}}\int (K^{''}(u))^{2}\left( f(x)-uhf^{'}(x)+O(h^{2})\right) du-\frac{1}{n}\left( f^{''}(x)\right)^{2}+ O \left(\frac{h^{2}}{n}\right)\\
  &=&\frac{1}{nh^{5}}f(x)\int (K^{''}(u))^{2}du+O\left( (nh^{4})^{-1}\right).
\end{eqnarray*}
Therefore
\begin{eqnarray}\label{3}
I =\frac{1}{nh^{5}}f^{''}(x)\int (K^{''}(u))^{2}du+O\left( (nh^{4})^{-1}\right).
\end{eqnarray}
From (\ref{varR}), (\ref{3}) and (\ref{varr}), we get (\ref{V}).\\

Now we consider
\begin{eqnarray*}
 MSE(\hat{f}_{n,h}^{b}(x))= Bias^{2}(\hat{f}_{n,h}^{b}(x))+ var(\hat{f}_{n,h}^{b}(x)).
\end{eqnarray*}
In our case, we have
\begin{eqnarray}\label{MMM} 
 MSE(\hat{f}_{n,h}^{b}(x)) \leq \frac{h^{6}}{36}\mu_{3}^{2}(f^{'''}(x))^{2}
                              + \frac{1}{2nh}\mu_{2}^{2}f(x) \int (K^{''}(u))^{2}du.                     
\end{eqnarray}
Minimizing the term on the right-hand side of this inequality yields $ h_{opt}=O(n^{-\frac{1}{7}}). $

 \end{pf}

 In general, there are many methods for selecting the practical bandwidth 
for the ordinary kernel density estimator:\\
\textsl{i}. One can experiment by using different bandwidths and simply select one 
that "looks right" for the type of data under investigation (subjective selection) \cite{JS}. \\
\textsl{ii}. One can refer to some given distribution, i.e. one selects the
 bandwidth that would be optimal for a particular pdf.\\
\textsl{iii}. One can use the cross-validation method introduced by Rudemo and Bowman \cite{TB} which provides 
an optimal bandwidth defined by
\begin{eqnarray}\label{hCV}
 h_{CV}=arg\min_{h>0}CV(h)
\end{eqnarray}
where $CV(h)$ is cross-validation given by $ CV(h)=\int \hat{f}^{2}dx-\frac{2}{n}\sum_{i=1}^{n}\hat{f},_{-i}(X_{i}) $ 
and $\hat{f},_{-i}(x)=\frac{1}{(n-1)h}{ \sum_{\stackrel{j=1 }{ j\neq i } }^{n} } K\left( \frac{x-X_{j}}{h}\right)$.\\
Following this idea, we propose a cross-validation bandwidth selection for the bias reduced kernel density estimator

\section{Cross-validation bandwidth selection for the bias reduced kernel density estimator } 
\label{Cross}
The expression of Mean Integrated Squared  Error (MISE) is defined by
\begin{eqnarray*}
 MISE(\hat{f}_{n,h}^{b}(x)):= \int \left( Bias^{2}(\hat{f}_{n,h}^{b}(x))+ Var(\hat{f}_{n,h}^{b}(x)) \right) dx.
\end{eqnarray*}
 Write $ MISE^{b}=MISE^{b}(h) $ to indicate that the mean integrated squared error is a function of bandwidth.
First, note that 
\begin{eqnarray*}
MISE (h)^{b} & = & \mathbb{E}_{f} \int(\hat{f}_{n,h}^{b}(x)-f(x))^{2}dx\\ \nonumber
 &=& \mathbb{E}_{f} \left[\int \left( \hat{f}_{n,h}^{b}(x)\right)^{2}dx-2\int \hat{f}_{n,h}^{b} f(x) dx\right]+\int f^{2}(x)dx. 
\end{eqnarray*} 
 Since the integral $ \int f^{2}(x)dx $ does not depend on $ h, $ the minimizer of $MISE^{b}(h)$ also minimizes the function
 \begin{eqnarray*}
 J(h)= MISE^{b} (h)- \int f^{2}(x)dx = \mathbb{E}_{f} \left[\int \left( \hat{f}_{n,h}^{b}(x)\right)^{2}dx-2\int \hat{f}_{n,h}^{b}f(x)dx \right].
 \end{eqnarray*}
 Since $J(h)$ depends on the unknown density $f$, we rather use a Modified Cross-Validation estimator $MCV(h)$ of $J(h)$. 
 For this purpose, it is sufficient to consider the following estimators of each of the quantities $ \mathbb{E}_{f}\left[\int \left( \hat{f}_{n,h}^{b}(x)\right)^{2}dx\right]  $ 
 and $ \mathbb{E}_{f}\left[\int \hat{f}_{n,h}^{b} f(x)dx\right] $: 
  \begin{enumerate}
\item[$ \bullet $] An  unbiased estimator of $ \mathbb{E}_{f}\left[\int \left( \hat{f}_{n,h}^{b}(x)\right)^{2}dx\right]  $ is given by  $ \int \left( \hat{f}_{n,h}^{b}(x)\right)^{2}dx $.\\
\item[$ \bullet $] An unbiased estimator of $ \mathbb{E}_{f}\left[\int \hat{f}_{n,h}^{b} f(x)dx\right] $  is given by 
$\frac{1}{n}\sum_{i=1}^{n}\hat{f}_{n,h}^{b},_{-i}(X_{i})$
\end{enumerate}
where
 \begin{eqnarray*}
\hat{f}_{n,h}^{b},_{-i}(x)&=&\frac{1}{(n-1)h}{ \sum_{\stackrel{j=1 }{ j\neq i }}^{n} } K\left( \frac{x-X_{j}}{h}\right) -\frac{1}{2(n-1)h}\mu_{2}{ \sum_{\stackrel{j=1 }{ j\neq i }}^{n} } K^{''}\left( \frac{x-X_{j}}{h}\right).
\end{eqnarray*}
Consequently the estimator $MCV(h)$ is given by
\begin{eqnarray}\label{MCV}
MCV(h)=\int \left( \hat{f}_{n,h}^{b}(x)\right)^{2} dx-\frac{2}{n}\sum_{i=1}^{n}\hat{f}_{n,h}^{b},_{-i}(X_{i}),
\end{eqnarray} 
We deduce from (\ref{MCV}) the expression of optimal bandwidth $ h_{MCV}^{b} $  as follows
\begin{eqnarray}\label{hMCV}
h_{MCV}^{b}=arg\min_{h>0}MCV(h),
\end{eqnarray}
and the corresponding bias reduced kernel density estimator $\hat{f}^{b}_{n,h}$ of  $f$ is writen as:
\begin{eqnarray*}\label{E12}
\hat{f}_{n,h}^{b}(x)=\frac{1}{nh_{MCV}^{b}}\sum_{i=1}^{n}K\left(\frac{x-X_{i}}{h_{MCV}^{b}}\right)-\frac{1}{2nh_{MCV}^{b}}\mu_{2}\sum_{i=1}^{n}K^{''}\left(\frac{x-X_{i}}{h_{MCV}^{b}} \right).
\end{eqnarray*}
In this following section, we establish the strong consistency of the bias reduced kernel density estimator.

\section{Strong consistency of a bias reduced kernel density estimator}

 Let $X_{1}, X_{2},..., X_{n} $ be iid random variables of unknown density $f$. 
 Under some conditions on $f$ and  $K$, one obtains a strongly consistent estimator $ \hat{f}_{n,h}^{b}$ of $f$. 
 For proving such consistency results, we shall consider
 the following  regularity conditions. \\
 (H.1) $K$ is a density of a centered random variable with finite variance $\mu_{2}$. \\
(H.2) Set $\varphi=K-\frac{\mu_{2}}{2} K^{''}$; $ \left\| \varphi\right\|_{\infty} =\sup_{x\in \mathbb{R}} \left| \varphi (x)\right| :=\gamma < +\infty $
and $ \left\| \varphi\right\|_{2}:=\left( \int \varphi^{2}(u)du\right)^{1/2} < +\infty $. \\
 Consider the class of functions 
$\Phi=\left\lbrace t \mapsto \varphi\left((x-t)/h\right):h>0,~~x\in \mathbb{R} \right\rbrace $.
For $\varepsilon >0,$ let $N(\varepsilon, \Phi)=\sup_{Q}N(\gamma\varepsilon,\Phi, d_{2Q})$
where the supremum is taken over all probability measures $Q$ on $(\mathbb{R}, \mathcal{B}), d_{2Q}$ is the $L_{2}(Q)$-metric 
and  $N(\varepsilon, \Phi,d_{2Q} )$ is the minimal number of balls of radius $\varepsilon$ needed to cover $\Phi$. \\
(H.3) For some $C>0$ and $\nu >0,~N(\varepsilon, \Phi)\leq C\varepsilon^{-\nu}, 0<\varepsilon<1.$ \\
This condition discussed in \cite{JW, NP} holds 
whenever $ \varphi: \mathbb{R}\rightarrow \mathbb{R} $  is a function of bounded variation.\\
(H.4) $\Phi$ is a pointwise measurable class, that is, there exists a countable subclass $\Phi_{0} $ of $\Phi$ such that 
we can find for any function $\phi \in \Phi $ a sequence of functions $ {\phi _{m}}$ in $\Phi_{0} $ for
which $\phi_{m}(y)\rightarrow \phi(y),~~y\in \mathbb{R}$. \\
This condition is  satisfied whenever $\varphi$ is right
continuous.\\
(H.5) $f$ is four differentiable in neighbourhood of $x$.
 \begin{thm}
\label{theom}
 Assuming (H.1-H.5) are satisfied. For each pair of sequences  $ (a_{n})_{n\geq 1}$ and $ (b_{n})_{n\geq 1} $
 such that  $ \forall n\geq 1, 0< a_{n} < b_{n} \leq 1$ and $ a_{n} \leq h \leq b_{n}$ we have  with probability 1, 
\begin{eqnarray}
 \limsup_{n\rightarrow \infty} \sup_{a_{n} \leq h \leq b_{n}}\frac{\sqrt{nh} \left\| \hat{f}_{n,h}^{b}-\mathbb{E}\hat{f}_{n,h}^{b}\right\|_{\infty}}{\sqrt{\log \left( 1/h\right) \vee \log \log n } } =: \omega <\infty.
\end{eqnarray}
\end{thm}
The proof of this theorem follows along the lines of the proof of theorem 1 \cite{EMD} and requires the following two lemmas that provide some results on pointwise measurable class of bounded functions.\bigskip

Let $X_{i},~1\leq i \leq n $ be iid random variables  defined from a probability space $(\Omega, \mathcal{F},\mathbb{P})$ to a measurable space $(S, \mathcal{S})$.
Let $\mathcal{G}$ be a pointwise measurable class of bounded functions
and $G$ be  a finite-valued measurable function satisfying for all $x\in S $, $ G(x)\geq \sup_{g\in \mathcal{G}} \left| g(x)\right|$. 
Define  $\alpha_{n}$ to be the empirical process based on the sample $X_{1},...,X_{n}$, that is, if $g: S\rightarrow \mathbb{R},$ we have 
\begin{eqnarray*}
 \alpha_{n}(g)=\sum_{i=1}^{n} \left( g(X_{i}) - \mathbb{E}g(X) \right) / \sqrt{n},
\end{eqnarray*}
and set for any class $\mathcal{G}$ of such functions
\begin{eqnarray*}
 \left\| \sqrt{n} \alpha_{n} \right\|_{\mathcal{G}}=\sup_{g\in \mathcal{G}}\left| \sqrt{n} \alpha_{n} (g)\right|.
\end{eqnarray*}

\begin{lem}{(Corollary 4 \cite{EMD} )} Let $\mathcal{G}$ be a pointwise measurable class of bounded functions.
\label{lemm22}
If $C, \nu \geq 1 $ and $ 0 < \sigma \leq \beta $, the
following conditions hold  :\\
\begin{eqnarray*}
& & (1)~\mathbb{E}\left[ G(X)^{2} \right] \leq \beta^{2}\\
& &  (2)~N(\varepsilon, \mathcal{G}) \leq C\varepsilon^{-\nu}, ~~0<\varepsilon < 1\\
& &  (3)~\sigma_{0}^{2}:= \sup_{g\in \mathcal{G}}\mathbb{E}\left[ g(X)^{2} \right] \leq \sigma^{2} \\
& &  (4)~\sup_{g\in \mathcal{G}}\left\| g \right\|_{\infty} \leq U,~ where~~ \sigma_{0} \leq U \leq C_{2} \sqrt{n} \beta, 
 ~and~C_{2}=\frac{1}{4\sqrt{\nu \log C_{1}}},~C_{1}=C^{1/\nu}.  
\end{eqnarray*}
Then we have for some absolute constant $A$, 
\begin{eqnarray*}
 \mathbb{E}\left\| \sum_{i=1}^{n} \varepsilon_{i}g(X_{i})\right\|_{\mathcal{G}} \leq A\sqrt{\nu n \sigma_{0}^{2} \log \left( C_{1}\beta / \sigma_{0}\right) }
 + 2 A \nu U \log \left( C_{3} n \left( \beta / U \right)^{2}\right)~~with~~C_{3}=C_{1}^{2}/16\nu, 
\end{eqnarray*}
where $\varepsilon_{i},~1\leq i \leq n, $ is a sequence of
independent Rademacher random variables $X_{1},...,X_{n}$.
\end{lem}

\begin{lem}{(Inequality of Talagrand \cite{T})}
\label{INEQUALITY}
Let  $\mathcal{G}$   be a pointwise measurable class of functions satisfying
for some $0 < M < \infty $, 
\begin{eqnarray*}
 \left\| g \right\|_{\infty} \leq M, ~~~g\in \mathcal{G}.
\end{eqnarray*}
 Then we have for all $t > 0 $, 
\begin{eqnarray*}
 & & \mathbb{P} \left\lbrace  \max_{1 \leq m \leq n } \left\|  \sqrt{m}\alpha_{m}\right\|_{\mathcal{G}} \geq A_{1}\left( \mathbb{E}\left\| \sum_{i=1}^{n} \varepsilon_{i}g(X_{i})\right\|_{\mathcal{G}} +t \right) \right\rbrace\\
  & & \leq 2\left\lbrace \exp \left( -\frac{A_{2}t^{2}}{n\sigma_{\mathcal{G}}^{2}}\right) + \exp \left( -\frac{A_{2}t}{M} \right)\right\rbrace ,
\end{eqnarray*}
 where  $\sigma_{\mathcal{G}}^{2}=\sup_{g\in \mathcal{G}} Var (g(X)) $ and $A_{1}, A_{2}$ are universal constants.
\end{lem}

\textit{Proof of the theorem \ref{theom}.}\\
We first write that
\begin{eqnarray*}
 \mathbb{E}\left[ \varphi^{2} \left( \frac{x-X}{h}\right) \right] &=& h\int \frac{1}{h} \varphi^{2} \left( \frac{x-y}{h}\right)f(y)dy\\
                                                            & \leq & h \left\| f \right\|_{\infty} \int \varphi^{2}(u)du\\
                                                          & \leq & h  \left\| f \right\|_{\infty} \left\| \varphi \right\|_{2}^{2}. \\
\end{eqnarray*}
For $j,k \geq 0$ and $c> 0 $, set $ n_{k}=2^{k},~~h_{j,k}=\left( 2^{j}c \log n_{k}\right)/ n_{k}$ and 
\begin{eqnarray*}
 \Phi_{j,k}= \left\lbrace \varphi \left( \frac{x-.}{h}\right),~~h_{j,k} \leq h \leq h_{j+1,k},~x\in \mathbb{R} \right\rbrace.
\end{eqnarray*}

Therefore for $h_{j,k} \leq h \leq h_{j+1,k},$ one has 
\begin{eqnarray}\label{1}
  \mathbb{E}\left[ \varphi^{2} \left( \frac{x-X}{h}\right) \right] \leq  2 h_{j,k}  \left\| f \right\|_{\infty} \left\| \varphi \right\|_{2}^{2}.                                                          
\end{eqnarray}
On the other hand, using (H.2),
\begin{eqnarray}\label{2}
  \mathbb{E}\left[ \varphi^{2} \left( \frac{x-X}{h}\right) \right] \leq  \gamma^{2}.                                                          
\end{eqnarray}
Combining (\ref{1}) and (\ref{2}), we have
\begin{eqnarray*}
  \mathbb{E}\left[ \varphi^{2} \left( \frac{x-X}{h}\right) \right] &\leq &  \gamma^{2}  \wedge   2 h_{j,k}  \left\| f \right\|_{\infty} \left\| \varphi \right\|_{2}^{2} \\
                                                             & := & \gamma^{2}  \wedge B_{0}h_{j,k}:= \sigma_{j,k}^{2}~~where~~B_{0}= 2\left\| f \right\|_{\infty} \left\| \varphi \right\|_{2}^{2}~and~a \wedge b := \min(a,b).
\end{eqnarray*}
We now use the lemma \ref{lemm22} to bound
\begin{eqnarray*}
 \mathbb{E} \left\| \sum_{i=1}^{n_{k}} \varepsilon_{i}g(X_{i})\right\|_{\Phi_{j,k}}.
\end{eqnarray*}
We first note that each $ \Phi_{j,k} $ satisfies the condition 1 with $G=\beta= \gamma $. Further, since
$\Phi_{j,k} \subset \Phi $, we see by (H.3) that each $ \Phi_{j,k} $  also fulfills the condition 2. Without loss of generality  we assume
that $\nu, C \geq 1$ in (H.3). Noting that
\begin{eqnarray*}
 C_{1}\beta/ \sigma_{0} \leq \left( \beta^{2}/ \sigma_{0}^{2} \vee C_{1}^{2} \right)~~with~~a\vee b :=max(a,b).
\end{eqnarray*}
Applying lemma \ref{lemm22} with $ U = \beta = \gamma $ and using the bound $ \sigma_{0} \leq \sigma_{j,k} \leq \sqrt{B_{0}h_{j,k}} $,
 we have for $j \geq 0 $,
\begin{eqnarray*}
 \mathbb{E} \left\| \sum_{i=1}^{n_{k}} \varepsilon_{i}g(X_{i})\right\|_{\Phi_{j,k}}\leq A\sqrt{\nu n_{k}B_{0}h_{j,k} \log \left( \beta^{2}/ \sigma_{0}^{2} \vee C_{1}^{2}\right) }
 + 2 A \nu \gamma \log \left( C_{3} n_{k} \right)
\end{eqnarray*}
which is writen for $B_{1}=A\sqrt{\nu B_{0}}$ and $ B_{2}= B_{0}/\beta^{2}$  as
\begin{eqnarray*}
 \mathbb{E} \left\| \sum_{i=1}^{n_{k}} \varepsilon_{i}g(X_{i})\right\|_{\Phi_{j,k}} \leq B_{1}\sqrt{ n_{k}h_{j,k} \log \left( \frac{1}{B_{2}h_{j,k}} \vee C_{1}^{2}\right) }
 + 2 A \nu \gamma \log \left( C_{3} n_{k} \right).
\end{eqnarray*}
Using the fact that $ h_{j,k}\geq \left(c \log n_{k}\right)/ n_{k} $, for large $k$,
\begin{eqnarray*}
 B_{1}\sqrt{c\log n_{k}} \sqrt{\log \left( \frac{n_{k}}{c B_{2}\log n_{k}}\right)} \leq B_{1}\sqrt{ n_{k}h_{j,k} \log \left( \frac{1}{B_{2}h_{j,k}}  \vee C_{1}^{2} \right) }.
\end{eqnarray*}
And  for large $k$,
\begin{eqnarray*}
  \frac{2 A \nu \gamma \log \left( C_{3} n_{k} \right)}{B_{1}\sqrt{c\log n_{k}} \sqrt{\log \left( \frac{n_{k}}{cB_{2}\log n_{k}}\right)} } \leq B^{'}
\end{eqnarray*}
where $B^{'}=\frac{2 A \nu \gamma}{B_{1}\sqrt{c} }$. This implies that
\begin{eqnarray*}
2 A \nu \gamma \log \left( C_{3} n_{k} \right) \leq  B^{'} B_{1}\sqrt{c\log n_{k}} \sqrt{\log \left( \frac{n_{k}}{cB_{2}\log n_{k}} \right) } \leq B^{'} B_{1}\sqrt{ n_{k}h_{j,k} \log \left(\frac{1}{B_{2}h_{j,k}} \vee C_{1}^{2}\right)}.
\end{eqnarray*}
Therefore 
\begin{eqnarray*}
 \mathbb{E} \left\| \sum_{i=1}^{n_{k}} \varepsilon_{i}g(X_{i})\right\|_{\Phi_{j,k}} &\leq &  B_{1}\sqrt{ n_{k}h_{j,k} \log \left(\frac{1}{B_{2}h_{j,k}} \vee C_{1}^{2}\right) }
 + B^{'} B_{1}\sqrt{ n_{k}h_{j,k} \log \left( \frac{1}{B_{2}h_{j,k}} \vee C_{1}^{2}\right) }\\
                            &\leq & B_{3}\sqrt{ n_{k} h_{j,k} \log \left( \frac{1}{B_{2} h_{j,k}} \vee \log \log n_{k}\right) }\\
                           &:= & B_{3} a_{j,k}
\end{eqnarray*}
where $B_{3}= (B_{1}+B^{'} B_{1}) $. \\
Using  the lemma \ref{INEQUALITY} with $M=\gamma $ and $ \sigma_{\mathcal{G}}^{2}=\sigma_{\Phi_{j,k}}^{2}\leq B_{0}h_{j,k},$ 
we get for any $t>0,$
\begin{eqnarray*}
 & & \mathbb{P} \left\lbrace  \max_{n_{k-1} \leq n \leq n_{k} } \left\|  \sqrt{n}\alpha_{n}\right\|_{\Phi_{j,k}} \geq A_{1}\left(  B_{3} a_{j,k} +t \right) \right\rbrace
   \leq 2\left\lbrace \exp \left( -\frac{A_{2}t^{2}}{n_{j,k}B_{0}h_{j,k} }\right) + \exp \left( -\frac{A_{2}t}{K} \right)\right\rbrace.
\end{eqnarray*}
Setting for any $ \delta >1 $, with $t=\delta a_{j,k} $,~ $ j\geq 0$ and $k \geq 1$,
\begin{eqnarray*}
 p_{j,k}(\delta) =\mathbb{P} \left\lbrace  \max_{n_{k-1} \leq n \leq n_{k} } \left\|  \sqrt{n}\alpha_{n}\right\|_{\Phi_{j,k}} \geq A_{1}\left(  B_{3}  +\delta \right)a_{j,k} \right\rbrace
\end{eqnarray*}
and using the fact that $ \frac{a_{j,k}^{2}}{n_{k}h_{j,k} }\geq \log \log n_{k} $, we can infer that for large $k$,
\begin{eqnarray}\label{pj}
 p_{j,k}(\delta)   &\leq & 2\left\lbrace \exp \left( -\frac{A_{2}\delta^{2} a_{j,k}^{2}}{n_{j,k}B_{0}h_{j,k} }\right) + \exp \left( -\frac{A_{2}\delta a_{j,k}}{K} \right)\right\rbrace\\ \nonumber
                & \leq & 2\left\lbrace \exp \left( -\frac{A_{2}\delta^{2}}{B_{0} }\log \log n_{k} \right) + \exp \left( -\frac{A_{2}\delta }{K} \sqrt{n_{k}h_{j,k}   \log \log n_{k}} \right)\right\rbrace\\ \nonumber
                 & \leq & 2 \left( \log n_{k}\right)^{-\tilde{\delta} }~~where~~\tilde{\delta}=\frac{A_{2}}{B_{0}}\delta^{2}.
\end{eqnarray}
Set $ l_{k}=\max \left\lbrace j:h_{j,k}\leq 2 \right\rbrace $. For large $ k $
\begin{eqnarray}\label{lk}
 l_{k}\leq 2\log n_{k}.
\end{eqnarray}
Hence (\ref{pj}) and (\ref{lk}) give for large $k$ and for $\delta \geq 1 $
\begin{eqnarray*}
 P_{k}(\delta) & := & \sum_{j=0}^{l_{k}-1} p_{j,k}(\delta)  \leq  4 \left( \log n_{k} \right)^{1-\tilde{\delta}}
\end{eqnarray*}
which implies that if we choose $\delta \geq 2(B_{0}/A_{2})^{1/2}$, we have 
\begin{eqnarray}\label{5}
  P_{k}(\delta)  \leq \frac{4}{k^{3}(\log 2)^{3}}~~\textsl{and}~~
\sum_{k=1}^{\infty} P_{k}(\delta)  < \infty.
\end{eqnarray}
Notice from \cite{EMD} that  by definition of $l_{k} $ for large $k,~~h_{j,k}\leq 2 \Rightarrow 2h_{j,k} \geq 2~~and~~h_{j,k}\geq 1 $.
Consequently, we then have for $ n_{k-1} \leq n \leq n_{k} $, 
\begin{eqnarray*}
 \left[ \frac{c\log n}{n},1 \right] \subset \left[ \frac{c\log n_{k}}{n_{k}},h_{j,k} \right].
\end{eqnarray*}
Thus for all large enough $k$ and $ n_{k-1} \leq n \leq n_{k} $,
\begin{eqnarray*}
 A_{k}(\delta) & := & \left\lbrace \max_{n_{k-1} \leq n \leq n_{k}} \sup_{(c\log n)/n \leq h \leq 1} \frac{\sqrt{nh} \left\| \hat{f}_{n,h}^{b}-\mathbb{E}\hat{f}_{n,h}^{b}\right\|_{\infty}}{\sqrt{\log \left( 1/h\right) \vee \log \log n }} > 2A_{1}(B_{3}+\delta)\right\rbrace \\
             &  \subset & \cup_{j=0}^{l_{k}-1} \left\lbrace \max_{n_{k-1} \leq n \leq n_{k} } \left\|  \sqrt{n}\alpha_{n}\right\|_{\Phi_{j,k}} \geq A_{1}\left(  B_{3}  +\delta \right)a_{j,k} \right\rbrace.             
\end{eqnarray*}
Therefore
\begin{eqnarray*}
\mathbb{P} \left( A_{k}(\delta) \right) & := & \mathbb{P} \left\lbrace \max_{n_{k-1} \leq n \leq n_{k}} \sup_{(c\log n)/n \leq h \leq 1} \frac{\sqrt{nh} \left\| \hat{f}_{n,h}^{b}-\mathbb{E}\hat{f}_{n,h}^{b}\right\|_{\infty}}{\sqrt{\log \left( 1/h\right) \vee \log \log n }} > 2A_{1}(B_{3}+\delta)\right\rbrace \\
             &  \leq  & \sum_{j=0}^{l_{k}-1}  \mathbb{P} \left\lbrace \max_{n_{k-1} \leq n \leq n_{k} } \left\|  \sqrt{n}\alpha_{n}\right\|_{\Phi_{j,k}} \geq A_{1}\left(  B_{3}  +\delta \right)a_{j,k} \right\rbrace =P_{k}(\delta).            
\end{eqnarray*}
By (\ref{5}), $\sum_{k=1}^{\infty} \mathbb{P} \left( A_{k}(\delta) \right) <\infty$. 
Via the Borel-Cantelli lemma
\begin{eqnarray*}
\mathbb{P} \left\lbrace  \limsup_{n\rightarrow \infty} \max_{n_{k-1} \leq n \leq n_{k}} \sup_{(c\log n)/n \leq h \leq 1} \frac{\sqrt{nh} \left\| \hat{f}_{n,h}^{b}-\mathbb{E}\hat{f}_{n,h}^{b}\right\|_{\infty}}{\sqrt{\log \left( 1/h\right) \vee \log \log n }} > 2A_{1}(B_{3}+\delta)\right\rbrace =0.\\
\end{eqnarray*}
This implies that 
\begin{eqnarray*}
\mathbb{P} \left\lbrace \limsup_{n\rightarrow \infty}  \max_{n_{k-1} \leq n \leq n_{k}} \sup_{(c\log n)/n \leq h \leq 1} \frac{\sqrt{nh} \left\| \hat{f}_{n,h}^{b}-\mathbb{E}\hat{f}_{n,h}^{b}\right\|_{\infty}}{\sqrt{\log \left( 1/h\right) \vee \log \log n }} \leq  2A_{1}(B_{3}+\delta)\right\rbrace =1
\end{eqnarray*}
and
\begin{eqnarray}\label{B}
 \limsup_{n\rightarrow \infty}  \max_{n_{k-1} \leq n \leq n_{k}} \sup_{(c\log n)/n \leq h \leq 1} \frac{\sqrt{nh} \left\| \hat{f}_{n,h}^{b}-\mathbb{E}\hat{f}_{n,h}^{b}\right\|_{\infty}}{\sqrt{\log \left( 1/h\right) \vee \log \log n }} \leq  2A_{1}(B_{3}+\delta).
\end{eqnarray}
As $n\rightarrow \infty,~(c\log n)/n \rightarrow 0$.  From (\ref{B}), we can write for $0<h<1$ such that $a_{n} \leq h \leq b_{n}$ 
\begin{eqnarray*}
  \limsup_{n\rightarrow \infty}  \sup_{a_{n} \leq h \leq b_{n}} \frac{\sqrt{nh} \left\| \hat{f}_{n,h}^{b}-\mathbb{E}\hat{f}_{n,h}^{b}\right\|_{\infty}}{\sqrt{\log \left( 1/h\right) \vee \log \log n }} \leq  2A_{1}(B_{3}+\delta).
\end{eqnarray*}
\begin{rmk}
\label{rk2}
 We further note that Theorem \ref{theom} implies for any sequences $0 < a_{n} < b_{n} \leq 1$, satisfying
$ b_{n} \rightarrow 0$  and $ na_{n} / \log (n)\rightarrow \infty $, with probability 1,
\begin{eqnarray}\label{norm1}
\sup_{a_{n}\leq h \leq b_{n}} \left\| \hat{f}_{n,h}^{b}-\mathbb{E}\hat{f}_{n,h}^{b}\right\|_{\infty}= 0\left( \sqrt{\frac{\log (1/a_{n})\vee \log\log n}{n a_{n}}}\right);
\end{eqnarray}
which in turn implies 
\begin{eqnarray*}
 \lim_{n\rightarrow \infty } \sup_{a_{n}\leq h \leq b_{n}} \left\| \hat{f}_{n,h}^{b}-\mathbb{E}\hat{f}_{n,h}^{b}\right\|_{\infty} = 0~~~a.s.
\end{eqnarray*}
\end{rmk}

\begin{thm}
\label{theorem1}
 Let $ f $  be  Lipschitz  function on $\mathbb{R}$. Assume that  (H.1) and (H.5) are satisfied and 
  the derivative of order $j$  of $f$ are bounded, $\forall  j=2,3,4$.
For any sequences $0 < a_{n} < b_{n} \leq 1$ satisfying $ a_{n}\leq h \leq b_{n}$ together with $ b_{n} \rightarrow 0$, we have
 \begin{eqnarray*}\label{norm2}
\sup_{a_{n}\leq h \leq b_{n}} \left\| \mathbb{E}\hat{f}_{n,h}^{b}-f\right\|_{\infty} = 0\left( b_{n}\right).
\end{eqnarray*}
\end{thm}

\begin{pf}
 Set $ \Psi_{n,h}(x) = \mathbb{E}\hat{f}_{n,h}^{b}(x)-f(x),\forall x\in \mathbb{R} $. Using the formula (\ref{E1}) and (\ref{E2}), we have
 \begin{eqnarray*}
  \Psi_{n,h}(x)&=&   \mathbb{E} \hat{f}_{n,h}(x)-\frac{h^{2}}{2} \mu_{2} \mathbb{E}\hat{f}^{''}_{n,h}(x) -f(x)\\
            &=&  \mathbb{E} \left[ \frac{1}{nh}\sum_{i=1}^{n}K\left( \frac{x-X_{i}}{h}\right)\right]-
            \frac{h^{2}}{2}\mu_{2}\mathbb{E} \left[ \frac{1}{nh^{3}}\sum_{i=1}^{n}K^{''}\left( \frac{x-X_{i}}{h}\right)\right]-f(x)\\
            &=&\frac{1}{h}\int K\left( \frac{x-y}{h} \right) f(y)dy-\frac{h^{2}}{2}\mu_{2}\frac{1}{h^{3}}\int K^{''}\left( \frac{x-y}{h}\right)f(y)dy-f(x)\\                   
             &=& \frac{1}{h}\int K\left( \frac{x-y}{h} \right) f(y)dy-\frac{h^{2}}{2}\mu_{2}\frac{1}{h}\int K\left( \frac{x-y}{h}\right)f^{''}(y)dy-f(x) \\
             &=&  \int K(u)f(x-uh)du-\frac{h^{2}}{2}\mu_{2}\int K(u)f^{''}(x-uh)du-f(x).
 \end{eqnarray*}
Applying a Taylor approximation on the second term, we have
\begin{eqnarray*}
   \Psi_{n,h}(x)&=&  \int K(u)\left[ f(x-uh)-f(x)+f(x)\right] du+\\
   & & -\frac{h^{2}}{2}\mu_{2}\left[ \int K(u)\left(  f^{''}(x) -uhf^{'''}(x) + 
   \frac{1}{2} (uh)^{2}f^{''''}(x)+o(h^{2}) \right) \right] du-f(x) \\
               &=& \int K(u)\left[ f(x-uh)-f(x)\right] du-\frac{h^{2}}{2}\mu_{2} f^{''}(x) - \frac{h^{4}}{4}\mu_{2}^{2}f^{''''}(x) 
               + o(h^{4}).
 \end{eqnarray*}
 Since $f^{''}$ and  $f^{''''}$ are bounded on $\mathbb{R}$, i.e. $\forall x\in \mathbb{R}, $ there exists two constants $M$ and $N$ 
 such that $\left| f^{''}(x)\right| \leq M  $ and $\left| f^{''''}(x)\right| \leq N  $.
 We then have 
 \begin{eqnarray}\label{A}
 \left|  \Psi_{n,h}(x)\right| \leq  \int K(u)\left| f(x-uh)-f(x)\right| du+\left| \frac{h^{2}}{2}\mu_{2}M\right|+ \left| \frac{h^{4}}{4}\mu_{2}^{2}N \right|
               +\left| o(h^{4})\right|.
 \end{eqnarray}
 For small enough $ h $, (\ref{A}) gives
 \begin{eqnarray*}
  \left|  \Psi_{n,h}(x)\right| \leq  \int K(u)\left| f(x-uh)-f(x)\right| du.
 \end{eqnarray*}
 Note that $ f $ is  Lipschitz  function on $\mathbb{R}$ .i.e. for $\alpha>0 $ and for $x,y\in \mathbb{R}$,
  $ \left| f(x)-f(y)\right|\leq \alpha \left|x-y \right|$.
 Consequently 
 \begin{eqnarray*}
  \left|  \Psi_{n,h}(x)\right| &\leq & \int K(u)\alpha\left| uh\right| du\\
                 & \leq & \alpha\left|h\right|\int \left| u\right| K(u)du=  \alpha c \left|h\right|~~\textsl{where}~~c=\int \left| u\right| K(u)du<\infty.
 \end{eqnarray*}
 For any sequences $0 < a_{n} < b_{n} \leq 1$  satisfying $a_{n}\leq h \leq b_{n}$ together with $ b_{n} \rightarrow 0$, 
 we have
 \begin{eqnarray*}
  \left|  \Psi_{n,h}(x)\right| &\leq &   A \left|b_{n}\right|~~\textsl{with}~~A=\alpha c.
 \end{eqnarray*}
 Which means that 
 \begin{eqnarray*}
   \Psi_{n,h}(x) =  O(b_{n}).
 \end{eqnarray*}
 This finaly implies that 
 \begin{eqnarray*}
 \sup_{a_{n}\leq h\leq b_{n}} \left\| \Psi_{n,h}\right\|_{\infty} = O(b_{n}).
 \end{eqnarray*}
 It concludes the proof of the theorem.
\end{pf}

\section{ Strongly Consistent  Kullback-Leibler divergence Estimator }\label{sec2}

Let $X_{1},...,X_{n} $ be a random sample of unknown density function $f$ defined on $\mathbb{R}$
and  let $f_{\theta}$ be a parametric candidate model. 
Denote by $\mathcal{D}_{KL}(f,f_{\theta})$, the Kullback-Leibler divergence between $f$ and $f_{\theta}$ defined by
\begin{eqnarray}\label{con}
\mathcal{D}_{KL} (f,f_{\theta})=\int_{\mathbb{R}}f(x)\log{\left( \frac{f(x)}{f_{\theta}(x)} \right)}dx.
\end{eqnarray}
Notice that the Kullback-Leibler divergence  does not obey the triangle inequality  and
in general $\mathcal{D}_{KL} (f,f_{\theta})$ does not equal to $\mathcal{D}_{KL} (f_{\theta},f)$. The unknown density function $f$ can be
estimated by the bias reduced kernel density 
estimator. Using this estimator,  we then define the Kullback-Leibler divergence estimator of $\mathcal{D}_{KL}(f,f_{\theta})$  as follows
\begin{eqnarray}\label{estKL2}
 \widehat{\mathcal{D} }_{KL}(\hat{f}_{n,h}^{b},f_{\theta}):= \int_{A_{n}} \hat{f}_{n,h}^{b}(x) ln\left( \frac{ \hat{f}_{n,h}^{b}(x)}{f_{\theta}(x)}\right)  dx,
\end{eqnarray}
where $ A_{n}=\lbrace x\in \mathbb{R}; \hat{f}_{n,h}^{b}(x)~~ \geq \varepsilon_{n}\rbrace $ with
$ (\varepsilon_{n}) $  a sequence of positive constants such that $ \varepsilon_{n}\rightarrow 0 $ as $ n \rightarrow \infty $. 
Since $\hat{f}_{n,h}^{b}$  is strongly consistent as shown in preceding section,  we shall prove the strong consistency of 
Kulback-Leibler divergence estimator defined by (\ref{estKL2}). 
 Troughout the remainder of this paper $\hat{\mathbb{E}} \widehat{\mathcal{D} }_{KL}(\hat{f}_{n,h}^{b},f_{\theta}) $ is given by
\begin{eqnarray*}
\hat{\mathbb{E}} \widehat{\mathcal{D} }_{KL}(\hat{f}_{n,h}^{b},f_{\theta}):= \int_{A_{n}}\mathbb{E} \hat{f}_{n,h}^{b}(x) ln\left( \frac{\mathbb{E} \hat{f}_{n,h}^{b}(x)}{f_{\theta}(x)}\right)  dx,
\end{eqnarray*}
 where $ A_{n} $ is defined in (\ref{estKL2}). Hence the following theorem.
\begin{thm}
\label{theo1}
Let the conditions (H.1-H.5) be satisfied and  let $ f $ be  bounded and Lipschitz  density function on $\mathbb{R}$.   For each pair of sequence $(a_{n})_{n\geq 1}$ and $(b_{n})_{n\geq 1}$ and 
for $0<h<1$ such  that $ 0< a_{n} < h \leq b_{n} \leq 1$  together with $ b_{n}\rightarrow 0 $ and $ na_{n}/\log(n)\rightarrow \infty $  as $ n\rightarrow \infty $,
we have with probability $ 1 $;
\begin{eqnarray*}
\sup_{a_{n}\leq h\leq b_{n}} \left|  \widehat{\mathcal{D} }_{KL}(\hat{f}_{n,h}^{b},f_{\theta})- \mathcal{D}_{KL}(f,f_{\theta})\right| = O\left(\sqrt{\frac{\log (1/a_{n})\vee \log\log n}{n a_{n}}} \vee b_{n}\right).
\end{eqnarray*}
\end{thm}
The proof of this theorem is based on two following lemmas and  the methods developed in \cite{SBI} will be helpfull. 
\begin{lem}
\label{theo2}
Suppose that the conditions (H.1-H.5) hold and  let $ f $ be  continuous and bounded  density on $\mathbb{R} $. We have with probability $ 1 $, for each pair of
sequence $ 0<a_{n} < b_{n} \leq 1 $ and 
for $0<h<1$ such  that $ a_{n}\leq h \leq b_{n} $  together with  $ na_{n}/\log(n)\rightarrow \infty $  as $ n\rightarrow \infty $
\begin{eqnarray*}
\sup_{a_{n}\leq h\leq b_{n}}\left| \widehat{\mathcal{D} }_{KL}(\hat{f}_{n,h}^{b},f_{\theta})-\hat{\mathbb{E}} \widehat{\mathcal{D} }_{KL}(\hat{f}_{n,h}^{b},f_{\theta})\right| = O\left( \sqrt{\frac{\log (1/a_{n})\vee \log\log n}{n a_{n}}}\right).
\end{eqnarray*}
\end{lem}
We need the following  proposition in order to prove this lemma. 
\begin{prop}{ (Theorem 9.1. \cite{DL} ).}
\label{Devroye}
 Let $ K $ be an arbitrary integrable function on $ \mathbb{R}^{d} (i.e., \int \arrowvert K \arrowvert <\infty) $, and let $f$ be a
 density on $ \mathbb{R}^{d}$. Denoting $ K_{h}(x)=\left( 1/h^{d}\right) K\left( x/h\right), x\in \mathbb{R}^{d},h>0,$ we have 
 \begin{eqnarray*}
  \lim_{h \rightarrow 0} \int \left| f * K_{h}-f\int K \right| =0.
 \end{eqnarray*} 
\end{prop}

Proof of the lemma \ref{theo2}.\\ 
Define
\begin{eqnarray*}
\Gamma_{n1}:=\widehat{\mathcal{D} }_{KL}(\hat{f}_{n,h}^{b},f_{\theta})- \hat{\mathbb{E}} \widehat{\mathcal{D} }_{KL}(\hat{f}_{n,h}^{b},f_{\theta}).
\end{eqnarray*} 
One has 
\begin{eqnarray*}
\Gamma_{n1} &=& \int_{A_{n}} \left[ \hat{f}_{n,h}^{b}(x) ln\left( \hat{f}_{n,h}^{b}(x) \right) -\hat{f}_{n,h}^{b}(x) ln\left(f_{\theta}(x) \right)\right] dx - \int_{A_{n}}\left[  \mathbb{E} \hat{f}_{n,h}^{b}(x) ln \left(\mathbb{E}\hat{f}_{n,h}^{b}(x) \right) - \mathbb{E} \hat{f}_{n,h}^{b}(x)ln\left(f_{\theta}(x) \right)\right] dx \\
            &=& \int_{A_{n}} \left[ ln\left( \hat{f}_{n,h}^{b}(x) \right)-ln\left(\mathbb{E} \hat{f}_{n,h}^{b}(x) \right) \right]  \mathbb{E} \hat{f}_{n,h}^{b}(x)dx +\int_{A_{n}} \left[ \hat{f}_{n,h}^{b}(x)-\mathbb{E} \hat{f}_{n,h}^{b}(x)  \right] ln\left( \hat{f}_{n,h}^{b}(x) \right) dx +\\
            & & -  \int_{A_{n}} \left[ \hat{f}_{n,h}^{b}(x)- \mathbb{E} \hat{f}_{n,h}^{b}(x)  \right]  ln\left(f_{\theta}(x) \right)dx \\
             &:=& \Gamma_{n11} + \Gamma_{n12} -\Gamma_{n13}.             
 \end{eqnarray*} 
 We first prove that $ \sup_{a_{n}\leq h\leq b_{n}}\left| \Gamma_{n11} \right|= O\left( \sqrt{\frac{\log (1/a_{n})\vee \log\log n}{n a_{n}}}\right). $ Observing that for all $ y>0,\left| ln(y)\right| \leq \left| \frac{1}{y}-1 \right| + \left| y-1 \right| $, we have 
 \begin{eqnarray*}
 \left| ln\left( \hat{f}_{n,h}^{b}(x) \right)-ln\left(\mathbb{E} \hat{f}_{n,h}^{b}(x) \right)\right| &=& \left| ln  \frac{\hat{f}_{n,h}^{b}(x)}{\mathbb{E} \hat{f}_{n,h}^{b}(x)} \right| \\
      & \leq &     \frac{ \left| \mathbb{E} \hat{f}_{n,h}^{b}(x)-\hat{f}_{n,h}^{b}(x)\right| }{\hat{f}_{n,h}^{b}(x)}  +  \frac{\left| \hat{f}_{n,h}^{b}(x)-\mathbb{E} \hat{f}_{n,h}^{b}(x)\right|}{\mathbb{E} \hat{f}_{n,h}^{b}(x)}.  
 \end{eqnarray*}
 Recalling that $ A_{n}=\left\lbrace x\in \mathbb{R}, \hat{f}_{n,h}^{b}(x)\geq\varepsilon_{n} \right\rbrace  $, we readily obtain from these relations that, for any $ x\in A_{n}, $
 \begin{eqnarray*}
  \left| ln\left( \hat{f}_{n,h}^{b}(x) \right)-ln\left(\mathbb{E} \hat{f}_{n,h}^{b}(x) \right)\right| \leq \frac{2}{\varepsilon_{n}} \left| \hat{f}_{n,h}^{b}(x)- \mathbb{E} \hat{f}_{n,h}^{b}(x) \right|.
 \end{eqnarray*}
For any $ n\geq 1 $, we can therefore write the inegalities
 \begin{eqnarray*}
 \left| \Gamma_{n11} \right| &=& \left| \int_{A_{n}} \left[ ln\left( \hat{f}_{n,h}^{b}(x) \right)-ln\left(\mathbb{E} \hat{f}_{n,h}^{b}(x) \right)\right] \mathbb{E} \hat{f}_{n,h}^{b}(x) dx \right| \\
                         & \leq & \int_{A_{n}} \frac{2}{\varepsilon_{n}} \left| \hat{f}_{n,h}^{b}(x)- \mathbb{E} \hat{f}_{n,h}^{b}(x) \right| \mathbb{E} \hat{f}_{n,h}^{b}(x) dx \\                         
                         & \leq & \frac{2}{\varepsilon_{n}} \sup_{x\in A_{n}} \left| \hat{f}_{n,h}^{b}(x)- \mathbb{E} \hat{f}_{n,h}^{b}(x) \right| \int_{A_{n}} \mathbb{E} \hat{f}_{n,h}^{b}(x) dx\\
                         & \leq & \frac{2}{\varepsilon_{n}} \sup_{x \in \mathbb{R}} \left| \hat{f}_{n,h}^{b}(x)- \mathbb{E} \hat{f}_{n,h}^{b}(x) \right| \int_{\mathbb{R}} \mathbb{E} \hat{f}_{n,h}^{b}(x) dx.
 \end{eqnarray*} 
 An application of  proposition \ref{Devroye} gives
  \begin{eqnarray*}
  \lim_{h \rightarrow 0} \int \left|\mathbb{E}\hat{f}_{n,h}^{b}(x)-f(x)\int \left( K -\frac{\mu_{2}}{2}K^{''} \right)(x) dx \right|dx =0.
 \end{eqnarray*}
 This implies that 
 \begin{eqnarray*}
  \lim_{h \rightarrow 0} \int \mathbb{E}\hat{f}_{n,h}^{b}(x) dx =\int f(x) dx \int  \left( K -\frac{\mu_{2}}{2}K^{''} \right)(x)  dx.
 \end{eqnarray*} 
 Assuming that $\zeta:=\int \left( K -\frac{\mu_{2}}{2}K^{''} \right)(x) dx <\infty $, one has
 \begin{eqnarray*}
  \lim_{h \rightarrow 0} \int \mathbb{E}\hat{f}_{n,h}^{b}(x) dx = \zeta  \int f(x) dx.
 \end{eqnarray*} 
 Thus
\begin{eqnarray*}
 \left| \Gamma_{n11} \right| & \leq & \frac{2\zeta}{\varepsilon_{n}}\sup_{x \in \mathbb{R}} \left| \hat{f}_{n,h}^{b}(x)- \mathbb{E} \hat{f}_{n,h}^{b}(x) \right|  \\
                          & \leq & \frac{2\zeta}{\varepsilon_{n}} \parallel \hat{f}_{n,h}^{b}(x)- \mathbb{E} \hat{f}_{n,h}^{b}(x) \parallel_{\infty}.     
\end{eqnarray*} 
Therefore
\begin{eqnarray}\label{G11}
\sup_{a_{n}\leq h \leq b_{n}} \left| \Gamma_{n11}\right| = \frac{2\zeta}{\varepsilon_{n}} \sup_{a_{n}\leq h \leq b_{n}} \parallel \hat{f}_{n,h}^{b}(x)  - \mathbb{E} \hat{f}_{n,h}^{b}(x) \parallel_{\infty}. 
\end{eqnarray}  
 Substituting  (\ref{norm1}) in (\ref{G11}) the result follows.\\
We next prove that $  \sup_{a_{n}\leq h\leq b_{n}}\left| \Gamma_{n12} \right|=  O\left( \sqrt{\frac{\log (1/a_{n})\vee \log\log n}{n a_{n}}}\right)$. Since $ \left| ln(y) \right| \leq \frac{1}{y}+y $, for all $ y> 0 $,
one has
\begin{eqnarray*}
\left| \Gamma_{n12} \right| &=&  \left| \int_{A_{n}} \left[\hat{f}_{n,h}^{b}(x)-\mathbb{E}\hat{f}_{n,h}^{b}(x)  \right] ln\left(\hat{f}_{n,h}^{b}(x)  \right) dx \right|\\
             & \leq & \int_{A_{n}} \left| \hat{f}_{n,h}^{b}(x)-\mathbb{E}\hat{f}_{n,h}^{b}(x)  \right| \left[ \frac{1}{\hat{f}_{n,h}^{b}(x)} + \hat{f}_{n,h}^{b}(x)\right] dx.
\end{eqnarray*}
Similary as above, we get  for any $ x\in A_{n},  $
\begin{eqnarray*}
\frac{1}{\hat{f}_{n,h}^{b}(x)} + \hat{f}_{n,h}^{b}(x) &=& \left( \frac{1}{\hat{f}_{n,h}^{b}(x)\hat{f}_{n,h}^{b}(x)} +1 \right)\hat{f}_{n,h}^{b}(x)\\
            & \leq & \left(  \frac{1}{\varepsilon_{n}^{2}}+1  \right)\hat{f}_{n,h}^{b}(x). 
\end{eqnarray*}
Therefore, we have 
\begin{eqnarray*}
\left| \Gamma_{n12}\right| & \leq & \left(  \frac{1}{\varepsilon_{n}^{2}}+1  \right)  \int_{A_{n}} \left| \hat{f}_{n,h}^{b}(x)-\mathbb{E}\hat{f}_{n,h}^{b}(x) \right| \hat{f}_{n,h}^{b}(x) dx\\
                      & \leq &  \left(  \frac{1}{\varepsilon_{n}^{2}}+1  \right)  \sup_{x\in A_{n}} \left| \hat{f}_{n,h}^{b}(x)  - \mathbb{E} \hat{f}_{n,h}^{b}(x)\right| \int_{A_{n}} \hat{f}_{n,h}^{b}(x) dx\\
                      & \leq &  \left(  \frac{1}{\varepsilon_{n}^{2}}+1  \right)  \sup_{x\in \mathbb{R}} \left| \hat{f}_{n,h}^{b}(x)  - \mathbb{E} \hat{f}_{n,h}^{b}(x)\right| \int_{\mathbb{R}} \hat{f}_{n,h}^{b}(x) dx\\
                      & \leq & \left(  \frac{1}{\varepsilon_{n}^{2}}+1  \right)\zeta  \left\| \hat{f}_{n,h}^{b}(x)  - \mathbb{E} \hat{f}_{n,h}^{b}(x)\right\|_{\infty}.
\end{eqnarray*} 
Since $ a_{n} \leq h\leq b_{n} $ and $ b_{n}\rightarrow 0,$ as $ n\rightarrow \infty $ we have
\begin{eqnarray}\label{l12}
 \sup_{a_{n}\leq h \leq b_{n}} \left| \Gamma_{n12}\right| &\leq &  \left(  \frac{1}{\varepsilon_{n}^{2}}+1  \right)\zeta  \sup_{a_{n}\leq h \leq b_{n}} \parallel \hat{f}_{n,h}^{b}(x)  - \mathbb{E} \hat{f}_{n,h}^{b}(x)\parallel_{\infty}.
\end{eqnarray}  
 Substituting (\ref{norm1}) in (\ref{l12}) we have 
 \begin{eqnarray*}
\sup_{a_{n}\leq h \leq b_{n}} \left| \Gamma_{n12}\right| &=&   \left(  \frac{1}{\varepsilon_{n}^{2}}+1  \right)\zeta 0\left( \sqrt{\frac{\log (1/a_{n})\vee \log\log n}{n a_{n}}}\right).                  
\end{eqnarray*}  
Hence   
\begin{eqnarray}\label{G12}
 \sup_{a_{n}\leq h \leq b_{n}} \left| \Gamma_{n12}\right| =  0\left( \sqrt{\frac{\log (1/a_{n})\vee \log\log n}{n a_{n}}}\right).                  
\end{eqnarray}  
 We evaluate now the last term  $ \sup_{a_{n}\leq h\leq b_{n}}\left| \Gamma_{n13} \right|=0\left( \sqrt{\frac{\log (1/a_{n})\vee \log\log n}{n a_{n}}}\right).$ \\
 Consider
 \begin{eqnarray*}
\left| \Gamma_{n13} \right| &=& \left| \int_{A_{n}} \left[  \hat{f}_{n,h}^{b}(x) -\mathbb{E} \hat{f}_{n,h}^{b}(x)  \right]  ln\left(f_{\theta}(x) \right)dx\right| \\
                   & \leq &  \sup_{x\in A_{n}} \left| \hat{f}_{n,h}^{b}(x) -\mathbb{E} \hat{f}_{n,h}^{b}(x) \right|  \int_{A_{n}}  \left(\frac{1}{f_{\theta}(x)}+ f_{\theta}(x)\right)dx.
 \end{eqnarray*}
Therefore
\begin{eqnarray*}
\left| \Gamma_{n13} \right|  & \leq & \sup_{x\in \mathbb{R}} \left| \hat{f}_{n,h}^{b}(x) -\mathbb{E} \hat{f}_{n,h}^{b}(x) \right|  \int_{\mathbb{R}} \left(\frac{1}{f_{\theta}(x)}+ f_{\theta}(x)\right)dx.
\end{eqnarray*} 
For  $ a_{n}\leq h \leq b_{n} $ and $ b_{n}\rightarrow 0 $ as $ n\rightarrow \infty $ there exists a constant $ C_{1}= \int_{\mathbb{R}}  \left(\frac{1}{f_{\theta}(x)}+ f_{\theta}(x)\right)dx < \infty $ such that  
 \begin{eqnarray*}
\sup_{a_{n}\leq h \leq b_{n}} \left| \Gamma_{n13} \right|  & \leq & C_{1}\sup_{a_{n}\leq h \leq b_{n}} \parallel \hat{f}_{n,h}^{b}(x) -\mathbb{E} \hat{f}_{n,h}^{b}(x) \parallel_{\infty}.
 \end{eqnarray*}
Thus in view of (\ref{norm1}), we get
 \begin{eqnarray}\label{G13}
 \sup_{a_{n}\leq h \leq b_{n}} \left| \Gamma_{n13}\right| =  0\left( \sqrt{\frac{\log (1/a_{n})\vee \log\log n}{n a_{n}}}\right).                  
\end{eqnarray} 
Finaly, the combination of  (\ref{G11}), (\ref{G12}) and (\ref{G13}) gives 
  \begin{eqnarray*}
 \sup_{a_{n}\leq h \leq b_{n}} \left| \Gamma_{n1} \right|  =  0\left( \sqrt{\frac{\log (1/a_{n})\vee \log\log n}{n a_{n}}}\right).                  
\end{eqnarray*} 
 It concludes the proof of the lemma \ref{theo2}.
 \begin{lem} 
 \label{theo3}
Assuming (A.1), (H.1) and (H.5) and  let $ f $ be   Lipschitz density function on $\mathbb{R}$.  For each pair of sequence $ 0<a_{n}<b_{n}\leq 1 $ and 
for $0<h<1$ such  that $ a_{n}\leq h \leq b_{n} $  together with $ b_{n}\rightarrow 0 $   as $ n\rightarrow \infty $,
we have with probability $ 1 $;
\begin{eqnarray*}
\sup_{a_{n}\leq h\leq b_{n}}\left| \hat{\mathbb{E}} \widehat{\mathcal{D} }_{KL}(\hat{f}_{n,h}^{b},f_{\theta})- \mathcal{D}(f,f_{\theta})\right| = O\left( b_{n}\right).
\end{eqnarray*}
\end{lem}

\begin{pf}
We set $ \Gamma_{n2}=\hat{ \mathbb{E}} \widehat{\mathcal{D} }_{KL}(\hat{f}_{n,h}^{b},f_{\theta})- \mathcal{D}(f,f_{\theta}) $, therefore
\begin{eqnarray*}
 \Gamma_{n2}&=& \int_{A_{n}} \left[ \mathbb{E} \hat{f}_{n,h}^{b}(x) ln \left(\hat{f}_{n,h}^{b}(x) \right) \hat{f}_{n,h}^{b}(x)-\mathbb{E}\hat{f}_{n,h}^{b}(x) ln\left(f_{\theta}(x) \right)-f(x)ln(f(x))+ f(x)ln\left(f_{\theta}(x) \right)\right] dx\\
            &=& \int_{A_{n}} \mathbb{E} \hat{f}_{n,h}^{b}(x) \left[ ln \left(\mathbb{E}\hat{f}_{n,h}^{b}(x) \right)-ln(f(x))\right] dx+ \int_{A_{n}} ln(f(x)) \left[  \mathbb{E} \hat{f}_{n,h}^{b}(x)-f(x)\right]dx +\\
            & & - \int_{A_{n}}  \left[  \mathbb{E} \hat{f}_{n,h}^{b}(x)-f(x)\right]ln\left(f_{\theta}(x) \right)dx\\
            &:=& \Gamma_{n21} + \Gamma_{n22} -\Gamma_{n23}.
\end{eqnarray*}
Our purpose is to show that $  \sup_{a_{n}\leq h\leq b_{n}} \left| \Gamma_{n21} \right| =O(b_{n}),  \sup_{a_{n}\leq h\leq b_{n}}\left| \Gamma_{n22} \right| =O(b_{n})$ and $ \sup_{a_{n}\leq h\leq b_{n}}\mid \Gamma_{n23} \mid=O(b_{n})$.
Begin by the first term. We can write
\begin{eqnarray*}
\left| \Gamma_{n21}\right| &=& \left|\int_{A_{n}}  \left[ ln \left(\mathbb{E}\hat{f}_{n,h}^{b}(x) \right)-ln(f(x))\right]\mathbb{E}\hat{f}_{n,h}^{b}(x) dx\right|\\
                    & \leq &  \int_{A_{n}} \left| ln \left(\mathbb{E}\hat{f}_{n,h}^{b}(x) \right)-ln(f(x))\right| \mathbb{E}\hat{f}_{n,h}^{b}(x) dx.
\end{eqnarray*}
Repeating the arguments above in the terms $ \left| \Gamma_{n11} \right| $ with the formal change of $ \hat{f}_{n,h}^{b} $ by $ f$, one has
\begin{eqnarray*}
\left| \Gamma_{n11} \right| \leq  \sup_{x\in A_{n}}\left| \mathbb{E}\hat{f}_{n,h}^{b}(x)-f(x) \right| \int_{A_{n}}\left( \frac{1}{\varepsilon_{n}} + \frac{1}{f(x)}\right)dx. 
\end{eqnarray*}
There exists a constant $ C_{2}= \int_{A_{n}}\left( \frac{1}{\varepsilon_{n}} + \frac{1}{f(x)}\right)dx < \infty $ such that 
\begin{eqnarray}\label{Gn11}
\mid \Gamma_{n11} \mid & \leq &  C_{2}\sup_{x\in \mathbb{R}}\mid \mathbb{E}\hat{f}_{n,h}^{b}(x)-f(x) \mid \\ \nonumber
                       & \leq &  C_{2}\parallel \mathbb{E}\hat{f}_{n,h}^{b}(x)-f(x) \parallel_{\infty}. 
\end{eqnarray}
Applying the theorem \ref{theorem1}, 
we have for each $ a_{n}\leq h \leq b_{n},$ as $n \rightarrow \infty, $ 
\begin{eqnarray}\label{norm22}
 \sup_{a_{n}\leq h \leq b_{n}}  \left\| \mathbb{E}\hat{f}_{n,h}^{b}-f \right\|_{\infty} = 0(b_{n}).
\end{eqnarray}
(\ref{Gn11}) combined with (\ref{norm22}) gives
\begin{eqnarray*}
\sup_{a_{n}\leq h \leq b_{n}} \left| \Gamma_{n11} \right| \leq C_{2} O(b_{n}). 
\end{eqnarray*}
Finaly
\begin{eqnarray}\label{G21}
\sup_{a_{n}\leq h \leq b_{n}} \left| \Gamma_{n11} \right| = O(b_{n}). 
\end{eqnarray}
Now we can prove the second term $  \sup_{a_{n}\leq h\leq b_{n}}\left| \Gamma_{n22} \right| = O(b_{n}).  $\\
Since $ \left| ln(y) \right| \leq \frac{1}{y}+y, $ for all $ y> 0 $
\begin{eqnarray*}
\left| \Gamma_{n22} \right|  & = &  \left| \int_{A_{n}}  \mathbb{E} \left[ \hat{f}_{n,h}^{b}(x)-f(x) \right] ln(f(x)) dx \right|\\ 
                       & \leq & \int_{A_{n}} \left|  \mathbb{E} \hat{f}_{n,h}^{b}(x)-f(x)\right| \left(  \frac{1}{f(x)}+f(x) \right)  dx \\
                        & \leq & \sup_{x\in A_{n}} \left|  \mathbb{E} \hat{f}_{n,h}^{b}(x)-f(x)\right| \int_{A_{n}} \left(  \frac{1}{f(x)}+f(x) \right)  dx \\ 
                         & \leq & \sup_{x\in \mathbb{R}} \left|  \mathbb{E} \hat{f}_{n,h}^{b}(x)-f(x)\right| \int_{\mathbb{R}} \left(  \frac{1}{f(x)}+f(x) \right)  dx.                        
\end{eqnarray*}
As assumed above, $ \int_{\mathbb{R}} \left(  \frac{1}{f(x)}+f(x) \right)  dx <\infty $, 
this in turn, implies that
\begin{eqnarray*}
\sup_{a_{n}\leq h \leq b_{n}} \left| \Gamma_{n22} \right| & \leq & C_{3}\sup_{a_{n}\leq h \leq b_{n}} \left\| \mathbb{E} \hat{f}_{n,h}^{b}(x)-f(x) \right\|_{\infty}. 
\end{eqnarray*}
where $C_{3}=\int_{\mathbb{R}} \left(  \frac{1}{f(x)}+f(x) \right)  dx <\infty$. In view of theorem \ref{theorem1}, one has
\begin{eqnarray}\label{G22}
\sup_{a_{n}\leq h \leq b_{n}} \left| \Gamma_{n22} \right| = O(b_{n}).
\end{eqnarray}
The third term is given by 
\begin{eqnarray*}
\Gamma_{n23}:=\int_{A_{n}}  \left[  \mathbb{E} \hat{f}_{n,h}^{b}(x)-f(x)\right]ln\left(f_{\theta}(x) \right)dx. 
\end{eqnarray*}
Repeat the argument in terms of $ \Gamma_{n22} $ with the formal change of $ f $ by $ f_{\theta} $ and considering the
constant\\
$ C_{4}= \int_{\mathbb{R}} \left(  \frac{1}{f_{\theta}(x)}+f_{\theta}(x) \right) dx< \infty $, one has
\begin{eqnarray*}
\left| \Gamma_{n23} \right| \leq C_{4} \left\| \mathbb{E} \hat{f}_{n,h}^{b}(x)-f(x) \right\|_{\infty}.
\end{eqnarray*}
 Thus
\begin{eqnarray}\label{G23}
\sup_{a_{n}\leq h \leq b_{n}} \left| \Gamma_{n23} \right| \leq  C_{4}\sup_{a_{n}\leq h \leq b_{n}} \left\| \mathbb{E} \hat{f}_{n,h}^{b}(x)-f(x) \right\|_{\infty} = 0(b_{n}).
\end{eqnarray}
Finaly combining (\ref{G21}), (\ref{G22}) and (\ref{G23}), the proof of lemma (\ref{theo3})  is deduced.
\end{pf}

\textit{Proof of theorem \ref{theo1}.}\\ 
The combination of the lemma \ref{theo2} and lemma \ref{theo3} concludes the proof of the Theorem \ref{theo1}.

\section{Applications for  Hypothesis Testing in Models Selection }

 Let  $ \left( \mathcal{X}, \beta_{\mathcal{X}},F \right)  $ be the statistical space,
 $\mathcal{X}=\lbrace x_{1},x_{2},...,x_{M_{0}}\rbrace, ~\forall M_{0}\geq 1 $;~ $ \beta_{\mathcal{X}} $ is the $  \sigma$-algebra
of all the sub-sets of $\mathcal{X}$ and $(\mathcal{X},\beta_{\mathcal{X}}) $, the measurable space.\\
Let  
\begin{eqnarray}\label{Simp}
\Lambda_{M_{0}}=\left\lbrace  F=(F_{1},..., F_{M_{0}})^{T}; \forall x\in \mathbb{R}~\textsl{and}~i=1,...,M_{0},~F_{i}(x) > 0~\textsl{and} ~ \sum_{i=1}^{M_{0}}F_{i}(x)=1 \right\rbrace 
\end{eqnarray}
be the simplex of distributions $ M_{0} $-vectors. It is the set of discrete distributions.
One can define the parametric family of models as follows
\begin{eqnarray}\label{ParModel}
\mathcal{F}=\left\lbrace  F_{\theta}= \left( F_{1}(.,\theta),...,F_{M_{0}}(., \theta)\right)^{T}: \theta \in \Theta \right\rbrace, 
\end{eqnarray}
where  $ \Theta \subset \mathbb{R}^{M_{0}} $. To be more explicit, 
suppose that  we are sampling from a distribution $F_{X}(x)$. Divide the range of the distribution into $M_{0}$ mutually exclusive and exhaustive classes, say $I_{1},...,I_{M_{0}}$. Each
class has a probability of containing the random variable $X, P(X\in I_{i}) := F_{i},~~
i = 1,...,M_{0}$ and  each sample value $x$ falls into exactly one of the intervals.

\subsection{ Goodness-of-fit test}
 The parametric family of models defined by (\ref{ParModel})  may or may not contain the true model. If $ \mathcal{F} $ contains 
the true model, then there exists a $ \theta_{0} \in \Theta $ such that $ F = F_{\theta_{0}}  $ and the model is said to be correctly specified. \\
We consider now the case when the model is not specified \textit{i.e.} $ H_{1}: F \neq F_{\theta_{0}}   $. 
Based on Kullback-Leibler divergence, this alternative hypothesis is writen as $ H_{1}: \mathcal{D}_{KL}(F,F_{\theta_{0}}) \neq 0$ where
\begin{eqnarray} \label{dis}
 \mathcal{D}_{KL}(F, F_{\theta_{0}})=\sum_{i=1}^{M_{0}} F_{i} \log \left( \frac{F_{i}}{F_{\theta_{0}},i}\right)~~with~~F_{\theta_{0},i}=F_{i}(.,\theta),~i=1,...,M_{0}
\end{eqnarray}
We must reject the null hypothesis iff
$\mathcal{D}_{KL}(F,F_{\theta_{0}}) > c $
where $c$ must be chosen for getting a level $ \alpha $ test. 
In general it is not possible to get the exact distribution of the statistic $\mathcal{D}_{KL}(F,F_{\theta_{0}}) $
and we must use its asymptotic distribution.
Notice that the estimator of $ \mathcal{D}_{KL}(F, F_{\theta_{0}}) $ given by (\ref{dis}) is defined as follows
\begin{eqnarray*} \label{dise}
 \widehat{\mathcal{D} }_{KL}(\hat{F}_{n,h}^{b},F_{\hat{\theta}})=\sum_{i=1}^{M_{0}} \hat{F}_{(n,h)i}^{b} \log \left( \frac{\hat{F}_{(n,h)i}^{b}}{F_{\hat{\theta}},i}\right)
\end{eqnarray*}
where $ \hat{F}_{n,h}^{b} $ is a bias reduced kernel density estimator of  $F$.
In the following theorem we present the  asymptotic distribution of $\mathcal{D}_{KL}(F, F_{\theta_{0}})$.\\
Let us  introduce the two important regularity assumptions:\\
-$ (J_{1}) $  Under the regularity conditions on the dominated model $ F_{\theta_{0}} $, 
the MLE is unique and asymptoticly normal under $ F_{\theta_{0}},~\forall \theta_{0} $
\begin{eqnarray*}
& & 1)F_{\hat{\theta}}\stackrel{as}{\longrightarrow} F_{\theta_{0}}~~\textit{when}~~n\rightarrow \infty\\
& & 2) \sqrt{n}(\hat{\theta}-\theta_{0})\stackrel{\mathcal{L}}{\longrightarrow} N(0, I(\theta_{0})^{-1})
\end{eqnarray*}
where $I(\theta_{0})$ is Fisher information and $ n\rightarrow \infty$.\\
-$ (J_{2}) $ There exists  $ \theta_{0}\in \Theta~~\textit{and}~~ \wedge^{\ast} = \left(
\begin{array}{cc}
\wedge_{11} & \wedge_{12}\\
\wedge_{21} & \wedge_{22}
\end{array}
\right)
$
, with $ \wedge_{12}=\wedge_{21} $   such that 
\begin{eqnarray*}
\sqrt{n}\left(
\begin{array}{cc}
\hat{F}_{n,h}^{b}-F\\
F_{\hat{\theta}}-F_{\theta_{0}}
\end{array}
\right)
\stackrel{\mathcal{L}}{\longrightarrow} N(0,\wedge^{\ast}).
\end{eqnarray*}
\begin{thm}
\label{thm5}
Let $ \mathcal{D}_{KL}(F, F_{\theta_{0}}) $ be the Kullback-Leibler divergence between $F$ and $F_{\theta_{0}}$ and 
let $\widehat{\mathcal{D} }_{KL}(\hat{F}_{n,h}^{b},F_{\hat{\theta}_{0}})$ its estimator. 
Under $ H_{1}:F_{\theta}\neq F  $ (we have omitted $0$ on $\theta $ ) and  assuming
that the conditions $ (J_{1})$ and $ (J_{2}) $  hold, one has:
\begin{eqnarray*}
\sqrt{n}\left[  \widehat{\mathcal{D} }_{KL}(\hat{F}_{n,h}^{b},F_{\hat{\theta}})- \mathcal{D}_{KL}(F,F_{\theta}) \right]  \stackrel{\mathcal{L}}{\longrightarrow} N(0,\wedge_{\phi}^{2})
\end{eqnarray*}
where 
\begin{eqnarray}\label{varia}
\wedge_{\phi}^{2}=A^{T}\wedge_{11}A+ A^{T}\wedge_{12}W+ W^{T}\wedge_{12}W+W^{T}\wedge_{22}W,
\end{eqnarray}
$ A^{T}=(a_{1},...,a_{M_{0}}) $ is the vector of partial derivatives with respect to the components of the first variable with
\begin{eqnarray*}
a_{i}=\left(\frac{\partial}{\partial F_{i}}  \mathcal{D}_{KL}(F,F_{\theta})\right),\qquad i=1,...,M_{0} 
\end{eqnarray*}
 and $ W^{T}=(w_{1},...,w_{M_{0}}) $ is the vector of partial derivatives with respect to the components of the second variable with
\begin{eqnarray*}
w_{i}=\left(\frac{\partial}{\partial F_{\theta,i}}  \mathcal{D}_{KL}(F,F_{\theta})\right), \qquad i=1,...,M_{0};~~ F_{\theta,i}=F_{i}(.,\theta).
\end{eqnarray*}
\end{thm}
\begin{pf} A first order Taylor expansion gives
\begin{eqnarray*}
\widehat{\mathcal{D} }_{KL}(\hat{F}_{n,h}^{b},F_{\hat{\theta}})= \mathcal{D}_{KL}(F,F_{\theta})+A^{T}(\hat{F}_{n,h}^{b}-F)+ W^{T}(F_{\hat{\theta}}-F_{\theta})+o\left( \left\| F_{n,h}^{b}-F\right|+\left\| F_{\hat{\theta}}-F_{\theta}\right|\right).
\end{eqnarray*}
We observe that 
\begin{eqnarray*}
\sqrt{n}\left[ \widehat{\mathcal{D} }_{KL}(\hat{F}_{n,h}^{b},F_{\hat{\theta}})- \mathcal{D}_{KL}(F,F_{\theta})\right] =\sqrt{n}\left[  A^{T}(\hat{F}_{n,h}^{b}-F)+ W^{T}(F_{\hat{\theta}}-F_{\theta})\right] +\sqrt{n}  o\left(\left| \hat{F}_{n,h}^{b}-F\right|+\left| F_{\hat{\theta}}-F_{\theta}\right|\right).
\end{eqnarray*}
Since 
$ \sqrt{n} (F_{\hat{\theta}}-F_{\theta}) \stackrel{\mathcal{L}}{\longrightarrow}    N(0,\Sigma_{F_{\theta}} )$, when $ n\rightarrow \infty $, with $ \Sigma_{F_{\theta}}=diag(F_{\theta})-F_{\theta}F_{\theta}^{t} $; then $  \left| F_{\hat{\theta}}-F_{\theta} \right| = O_{p}\left( n^{-1/2}\right)$ and  $ \sqrt{n}  o\left\| F_{\hat{\theta}}-F_{\theta} \right\| = o_{p}\left( 1\right)$. Therefore $ \sqrt{n} o\left(\left| \hat{F}_{n,h}^{b}-F \right| + \left| F_{\hat{\theta}}-F_{\theta} \right| \right) =o_{p}(1). $\\
Hence
\begin{eqnarray*}
\sqrt{n}\left[ \widehat{\mathcal{D} }_{KL}(\hat{F}_{n,h}^{b},F_{\hat{\theta}})- \mathcal{D}_{KL}(F,F_{\theta})\right] =\sqrt{n}\left[  A^{T}(\hat{F}_{n,h}^{b}-F)+ W^{T}(F_{\hat{\theta}}-F_{\theta})\right] +  o_{p}(1).
\end{eqnarray*}
The random variables $ \sqrt{n}\left[ \widehat{\mathcal{D} }_{KL}(\hat{F}_{n,h}^{b},F_{\hat{\theta}})- \mathcal{D}_{KL}(F,F_{\theta})\right] $ and $ \sqrt{n}\left[  A^{T}(\hat{F}_{n,h}^{b}-F)+ W^{T}(F_{\hat{\theta}}-F_{\theta})\right] $ have the same asymptotic distribution. 
In view of $ J_{1} $ and $ J_{2} $ we have 
\begin{eqnarray*} 
\sqrt{n}\left[  A^{T}(\hat{F}_{n,h}^{b}-F)+ W^{T}(F_{\hat{\theta}}-F_{\theta})\right]\stackrel{\mathcal{L}}{\longrightarrow} N(0,\wedge_{\phi}^{2})
\end{eqnarray*} 
where $ \wedge_{\phi}^{2} $ is given by (\ref{varia}).
This completes the proof.
\end{pf}
It is possible to choose the model among a collection of candidate models which is close to the true distribution according 
to the Kullback-Leibler divergence thanks to the goodness-of-fit test. 

\subsection{Test for model selection based on Kullback-Leibler divergence}
\label{Test}
We propose now to take  two candidate parametric models $ F_{\theta_{1}} $ and $ F_{\theta_{2}} $, .i.e. $ F_{\theta_{1}} $ and $ F_{\theta_{2}}  \in  \mathcal{F}  $. 
For simplicity in the rest of the paper, we will note $\theta$ in place of $ \theta_{1} $  and $\gamma $ in place of $\theta_{2}$.
Based on Kullback-Leibler divergence; we would like to choose the  candidate model which is close to the true probability distribution $ F $; 
i.e. the minimized KLD. Our major work is to propose some tests for model selection as follows \\
$ H_{0}: \mathcal{D}_{KL}(F,F_{\theta})=\mathcal{D}_{KL}(F,F_{\gamma}) $ ~means that the two models are equivalent,\\
 $ H_{1}:\mathcal{D}_{KL}(F,F_{\theta}) \neq \mathcal{D}_{KL}(F,F_{\gamma})$ ~means that $ F_{\theta}$ is not equivalent to $ F_{\gamma} $. \\
   
To define the  model selection statistic, we consider 
\begin{eqnarray}\label{kappa}
\xi^{2}=(U_{\theta}-U_{\gamma}; S_{\theta}-S_{\gamma})^{T}\wedge^{\ast}(U_{\theta}-U_{\gamma}; S_{\theta}-S_{\gamma})
\end{eqnarray}
 the variance of 
\begin{eqnarray*}
\sqrt{n}(U_{\theta}-U_{\gamma}; S_{\theta}-S_{\gamma})^{T}\left(
\begin{array}{cc}
F_{n,h}-F\\
F_{\hat{\theta}}-F_{\theta}
\end{array}
\right).
\end{eqnarray*}
 where
$ U_{\theta}^{T}=(u_{1},...,u_{M_{0}}) $; with
\begin{eqnarray*}
u_{i}=\left(\frac{\partial}{\partial F_{i}}  \mathcal{D}_{KL}(F,F_{\theta})\right),\qquad i=1,...,M_{0},
\end{eqnarray*}
and
$ S_{\theta}^{T}=(s_{1},...,s_{M_{0}}) $; with
\begin{eqnarray*}
s_{i}=\left(\frac{\partial}{\partial F_{\theta, i}}  \mathcal{D}_{KL}(F,F_{\theta})\right),\qquad i=1,...,M_{0}.
\end{eqnarray*}

Since $ U_{\theta},U_{\gamma}, S_{\theta}, S_{\gamma} $ and $ \wedge^{\ast} $ are consistently estimated by  their sample analogues  $ U_{\hat{\theta}},U_{\hat{\gamma}}, S_{\hat{\theta}}, S_{\hat{\gamma}} $ and $ \hat{\wedge}^{\ast}. $
Hence $ \xi^{2} $ is consistently estimated by 
\begin{eqnarray*}
\hat{\xi}^{2}=(U_{\hat{\theta}}-U_{\hat{\gamma}}; S_{\hat{\theta}}-S_{\hat{\gamma}})^{T}\hat{\wedge}^{\ast}(U_{\hat{\theta}}-U_{\hat{\gamma}}; S_{\hat{\theta}}-S_{\hat{\gamma}}).
\end{eqnarray*}
Therefore we propose the model selection statistic  $ KL_{n} $ as follows
\begin{eqnarray}\label{U}
KL_{n}=\frac{\sqrt{n}}{\hat{\xi}}\left[ \widehat{\mathcal{D} }_{KL}(\hat{F}_{n,h}^{b},F_{\hat{\theta}})-\widehat{\mathcal{D} }_{KL}(\hat{F}_{n,h}^{b},F_{\hat{\gamma}}) \right].
\end{eqnarray}
It is possible to get the asymptotic distribution of $KL_{n}$. Hence the following theorem.
\begin{thm}
\label{thm7}
 (Asymptotic distribution of the $ KL_{n} $-statistic).\\
 Under the regularity assumptions $(J_{1})$ and $(J_{2})$, suppose that $ \xi \neq 0 $, then under the null hypothesis $ H_{0},~KL_{n}\longrightarrow N(0,1). $
\end{thm}
\begin{lem}\label{lemm3}
 Under the regularity assumptions $(J_{1})$ and $(J_{1})$, we have\\
\textit{(1) for the model $ F_{\theta} $}
\begin{eqnarray}
\widehat{\mathcal{D} }_{KL}(\hat{F}_{n,h}^{b},F_{\hat{\theta}})= \mathcal{D}_{KL}(F,F_{\theta})+U_{\theta}^{T}(\hat{F}_{n,h}^{b}-F)+ S_{\theta}^{T}(F_{\hat{\theta}}-F_{\theta})+o_{p}(1).
\end{eqnarray}
\textit{(2) for model $F_{\gamma}  $}
\begin{eqnarray}
\widehat{\mathcal{D} }_{KL}(\hat{F}_{n,h}^{b},F_{\hat{\gamma}})= \mathcal{D}_{KL}(F,F_{\gamma})+U_{\gamma}^{T}(\hat{F}_{n,h}^{b}-F)+ S_{\gamma}^{T}(F_{\hat{\gamma}}-F_{\gamma})+o_{p}(1).
\end{eqnarray}
\end{lem}
\begin{pf} 
The results follows from a first order Taylor expansion.
\end{pf}

Proof of theorem \ref{thm7}.\\
From the lemma \ref{lemm3}, it follows that 
\begin{eqnarray*}
 \widehat{\mathcal{D} }_{KL}(\hat{F}_{n,h}^{b},F_{\hat{\theta}})-\widehat{\mathcal{D} }_{KL}(\hat{F}_{n,h}^{b},F_{\hat{\gamma}})&=&\mathcal{D}_{KL}(F,F_{\theta})-\mathcal{D}_{KL}(F,F_{\gamma})+U_{\theta}^{T}(\hat{F}_{n,h}^{b}-F)-U_{\gamma}^{T}(\hat{F}_{n,h}^{b}-F)+\\
& & + S_{\theta}^{T}(F_{\hat{\theta}}-F_{\theta})-S_{\gamma}^{T}(F_{\hat{\gamma}}-F_{\gamma})+o_{p}(1).
\end{eqnarray*}
Under $ H_{0},~ \mathcal{D}_{KL}(F,F_{\theta})=\mathcal{D}_{KL}(F,F_{\gamma}), ~F_{\theta}=F_{\gamma} $ and $ F_{\hat{\theta}}=F_{\hat{\gamma}} $ we have
\begin{eqnarray*}
 \widehat{\mathcal{D} }_{KL}(\hat{F}_{n,h}^{b},F_{\hat{\theta}})-\widehat{\mathcal{D} }_{KL}(\hat{F}_{n,h}^{b},F_{\hat{\gamma}})&=& U_{\theta}^{T}(\hat{F}_{n,h}^{b}-F)-U_{\gamma}^{T}(\hat{F}_{n,h}^{b}-F)+S_{\theta}^{T}(F_{\hat{\theta}}-F_{\theta})-S_{\gamma}^{T}(F_{\hat{\gamma}}-F_{\gamma})+o_{p}(1)\\
&=& (U_{\theta}-U_{\gamma}, S_{\theta}-S_{\gamma})^{T}\left(
\begin{array}{cc}
\hat{F}_{n,h}^{b}-F\\
F_{\hat{\theta}}-F_{\theta}
\end{array}
\right)
+o_{p}(1).
\end{eqnarray*}
Finally, applying the Central Limit Theorem and assumptions $(J_{1})-(J_{2})$, one can get immediately $ KL_{n}\longrightarrow N(0,1). $
It concludes the proof of the theorem  \ref{thm7}.

\section{Simulation Study}

To illustrate the theory discussed in the preceding section, we consider the family of Poisson distribution and the family
of Geometric distribution. For more details, let us introduce them. The probability mass function (PMF) of a
Poisson distribution with the parameter $ \lambda $ is given by 
\begin{eqnarray*}
 p(x; \lambda )\equiv f_{P}(x) =\frac{e^{-\lambda}\lambda^{x}}{x!},~~x\in \mathbb{N}
\end{eqnarray*}
and the PMF of geometric distribution with the parameter $ \theta $ is given by
\begin{eqnarray*}
 g(x; \theta )\equiv f_{G}(x) =\theta (1-\theta)^{x-1}, ~~x\in \mathbb{N}.
\end{eqnarray*}
We consider various sets of experiments in which data are generated from the mixture of a Poisson and Geometric distributions. 
These two distributions are calibrated  so that their two means are close (9 and 10 respectively).
Hence the Data Generating Process (DGP) has density
\begin{eqnarray*}
 t(\pi)=\pi \textsl{Poisson}(9) + (1-\pi)\textsl{Geometric} (0.1)
 \end{eqnarray*}
 where $ \pi\in (0,1) $ is specific to each set of experiments. In each set, several random samples are drawn from this mixture. The sample size varies from $20$ to $250$, and for each sample size the number of replications is $1000$. 
We choose different values of $ \pi $ which are $ 0.00,0.25, 0.5, 0.75, 1.00. $  Although our proposed model selection procedure does not require that the data generating process belong
 to either of the candidate models. We consider the two limiting cases  $\pi= 0.00  $ and $= 1.00$ for they correspond to the correctly specified cases. For $ \pi=0.25 $ and $ \pi=0.75 $ both candidate models are misspecified but not at equal distance from the DGP. These cases correspond
to a DGP which is Poisson or Geometric distributions but slightly contaminated by the other distribution. The value $ \pi=0.5 $ is the value for which the Poisson and Geometric distributions are approximately
 at equal distance to the mixture $ t(\pi) $. In order to perfect fit by the proposed method, for the chosen parameters of these two distributions, 
 we note that most of the mass is concentrated between $0$ and $10$. Therefore, the chosen partition has eight cells defined by
$ \left\lbrace \left[ c_{i-1},c_{i} \right]= \left[i-1,i \right], i=1,...,7 \right\rbrace $ 
and $ \left[ c_{7},c_{8} \right]=\left[7, \infty \right] $.
The corresponding maximum likelihood estimators (MLEs) of  $ \lambda $ and $\theta $ are
$ \hat{\lambda}\equiv\bar{X}_{n}=\frac{1}{n}\sum_{i=1}^{n}X_{i}$
and $ \hat{\theta}=\frac{n}{n+\sum_{i=1}^{n}(x_{i}-1)} $. Since the properties of kernel density estimators do not depend much on which particular kernel is used, 
 we choose the standard normal as the kernel function $K$ without loss of generality. Therefore  for the Gaussian kernel,
\begin{eqnarray*}\label{fch} 
\hat{f}_{n,h}^{b}(x)= \frac{1}{2\sqrt{2\pi} nh}\sum_{i=1}^{n}\left[3- \left( \frac{x-X_{i}}{h}\right)^{2}\right]e^{-\frac{1}{2}\left( \frac{x-X_{i}}{h}\right)^{2}}.
\end{eqnarray*}
To get $ h $ optimal of a bias reduced kernel density estimator, the  cross-validation bandwidth selection for 
a bias reduced kernel density estimator is used as proved in section \ref{Cross}.

 \subsection{Comparative results of  $\hat{D}_{1}\equiv \widehat{\mathcal{D} }_{KL}(\hat{f}_{n,h},f_{\hat{\theta} })$ and $ \hat{D}_{2}\equiv \widehat{\mathcal{D} }_{KL}(\hat{f}_{n,h}^{b},f_{\hat{\theta} }) $.}

In this subsection, we carry out a simulation study designed to demonstrate the  performance of 
the Kullback-Leibler divergence estimator based on the
 bias reduced kernel density estimator $\hat{f}_{n,h}^{b}$ comparatively to the Kullback-Leibler divergence estimator based on the kernel density estimator $\hat{f}_{n,h}$ given in (\ref{E2}) and (\ref{E1}) respectively.
 To do this well, we use
 the Minimized Kullback-Leibler  Divergence (MKLD) defined by 
 $MKLD=\frac{\hat{D}_{2}}{\hat{D}_{1}}$ as a measure  of efficiency of the estimator,
 where $ f_{\theta}$ is a given parametric model.
 If this MKLD is less than 1, then we conclude that $\hat{f}_{n,h}^{b}$ is a more efficient
estimator of $f$ than $\hat{f}_{n,h}$ in this  sense that it has a smaller Kullback-Leibler divergence.
 In our case, we suppose  $f_{\theta}$  to be a Geometric distribution. The values in parentheses are standard errors.
 Note that to be too rigorous, we have used the classical cross-validation method in order to get $ h $ optimal when the kernel density estimator 
  is computed and the  cross-validation bandwidth selection for 
a bias reduced kernel density estimator when the bias reduced kernel density estimator is computed. The results are presented in \textbf{Tables 1-3}.\\

 \begin{center}
\textbf{ Table 1. DGP=  0.25 Poisson (9) + 0.75 Geometric (0.1)}

\begin{tabular}{lcccccr}
\hline
   n   & & 20 & 60  &   90   &   150  &  250   \\
 \hline
$ \hat{D}_{1} $ & & 0.3425  & 0.5857 & 0.7687 & 1.1369 & 1.8199 \\
                        & & ( 0.2968)  & (0.5109) &  ( 0.6302) & (0.1570)& (0.8641)\\
                         &   &           &          &          &             &          \\
$ \hat{D}_{2} $ & & 0.0894   &  0.3488 & 0.5500 &   0.9825 & 1.7547 \\
                        & &  (0.0421)  & (0.1082) & (0.1528) & (0.2370) & (0.3721)\\
                        & &           &          &          &                     \\                            
$ MKLD $ & & 0.2610    & 0.5955 & 0.7154  &   0.8641& 0.9641   \\
     &  & (0.2848)    & (0.3956) &  (0.4187)   & (0.3934) &  (0.3601)             \\                          
 \hline             
\end{tabular}
\end{center}

\begin{center}
\textbf{ Table 2. DGP=  0.5 Poisson (9) + 0.5 Geometric (0.1)}

\begin{tabular}{lcccccr}
\hline
   n   & & 20 & 60  &   90   &   150  &  250   \\
 \hline
$ \hat{D}_{1} $ & & 0.4865  & 1.0904 & 1.5158 &  2.4796 & 4.2242 \\
                        & & ( 0.3125)  & (0.5543) &  (0.6303) & (0.8219)& (1.1855)\\
                         &   &           &          &          &             &          \\
$ \hat{D}_{2} $ & & 0.0597   &  0.2283 & 0.3710 &   0.6707 & 1.2102 \\
               & &  ( 0.0295)  & (0.0669) & (0.0889) & (0.1374) & (0.2205)\\
               & &           &          &          &            &         \\                            
$ MKLD $ & & 0.1227    & 0.2093 & 0.2447  &   0.2704 & 0.2864   \\
         &  & ( 0.1196)    & (0.1272) &  (0.1093)   & (0.0942) &  (0.0873)             \\                          
 \hline             
\end{tabular}
\end{center}

\begin{center}
\textbf{ Table 3. DGP=  0.75 Poisson (9) + 0.25 Geometric (0.1)}

\begin{tabular}{lcccccr}
\hline
   n   & & 20 & 60  &   90   &   150  &  250   \\
 \hline
$ \hat{D}_{1} $ & & 0.6187  & 1.4410 & 2.0764 &  3.5116 & 6.0540 \\
                        & & ( 0.3373)  & (0.5570) &  (0.6995) & (0.9557)& (1.4543)\\
                         &   &           &          &          &             &          \\
$ \hat{D}_{2} $ & & 0.0511   &  0.1428 & 0.2270 &   0.4213 & 0.7760 \\
                & &  ( 0.0329)  & (0.0394) & (0.0523) & (0.0912) & (0.1405)\\
                & &           &          &          &                     \\                            
$ MKLD $ & & 0.0825    & 0.0990 &  0.1093  &   0.1199 & 0.1281   \\
     &  & (0.0520)    & (0.0519) &  (0.0477)   & (0.0402) &  (0.0334)             \\                          
 \hline             
\end{tabular}
\end{center}
All tables show that $ \hat{D}_{2} $ has a small values   comparatively to $ \hat{D}_{1}$. As consequence the minimized Kullback-Leibler divergence
(MKLD) is  less than 1. This proves that the Kullback-Leibler divergence estimator based on a bais reduced kernel density estimator 
 perform competitively well and
better than the Kullback-Leibler divergence estimator based on the ordinary kernel density estimator.

\subsection{Simulation results: Model selection procedure}
To illustrate the model selection procedure discussed in the subsection \ref{Test},  we consider the problem of
choosing between the family of Poisson distribution and the family of Geometric distribution.   
 The KLD between  the bias reduced kernel density  estimator and each model are defined as follows
\begin{eqnarray*} 
\widehat{\mathcal{D} }_{KL}(\hat{f}_{n,h}^{b}, f_{ \hat{P}})=\sum_{i=1}^{m} \hat{f}_{(n,h)i}^{b} \log \left( {\frac{\hat{f}_{(n,h)i}^{b}}{f_{\hat{P},i}}}\right)
\end{eqnarray*}
and
\begin{eqnarray*}
 \widehat{\mathcal{D} }_{KL}(\hat{f}_{n,h}^{b}, f_{ \hat{G}})=\sum_{i=1}^{m} \hat{f}_{(n,h)i}^{b} \log \left( {\frac{\hat{f}_{(n,h)i}^{b}}{f_{\hat{G},i}}} \right).
\end{eqnarray*}
where $m$ is the number of cells considered.
The results of our five sets of experiments are presented in \textbf{Tables 1-5}.

\begin{center}
\textbf{Table 1. DGP=Poisson (9)}  

\begin{tabular}{lcccccr}
\hline
   n   & & 20 & 60  &   90   &   150  &  250   \\
 \hline
$ \hat{\lambda} $ & & 8.9727 &    8.9817     & 8.9967  & 8.9946 &  8.9928 \\
                 & & (0.6482) &   (0.3858)  & (0.3194) &  (0.2331)& (0.1877)        \\
                 &   &           &          &          &             &          \\
$ \hat{\theta} $  & &  0.1006   &   0.1003   & 0.1001 &  0.1001 &  0.1001   \\
                & & (0.0065)    & (0.0039)&  (0.0031) &   ( 0.0023) & (0.0018)             \\
                 &   &           &          &          &             &          \\
$ \hat{ \mathcal{D}}_{KL_{1}} $ & & 0.1221  & 0.4186 & 0.6527 & 1.1472 & 1.9759 \\
                        & & ( 0.0339)  & (0.0707) &  ( 0.1000) & (0.1570)& (0.2549)\\
                         &   &           &          &          &             &          \\
$ \hat{ \mathcal{D}}_{KL_{2}} $ & &   0.1142   &  0.1626 & 0.2104 &  0.3060 & 0.4400 \\
                        & &  (0.0962)  & (0.1055) & (0.1231) & ( 0.1457) & (0.1836)\\
                        & &           &          &          &                     \\                            
$ KL_{n} $ & & 0.0774    & 2.3677 & 3.5413  &   5.0650 & 6.2434   \\
     &  & (0.4555)    & (0.8375) &  (1.1850)   & (2.0340) &  (3.8894)             \\                     
Model selection & Correct       & 4.6\%    & 72.4\%  & 94.8\% & 99.7\% & 100\% \\
based on $ KL_{n} $  & Indecisive  & 95.3\%  & 27.0\%  &  5.1\%  &  0.3\% & 0.0\%\\
                & Incorrect   &  0.1\% &   0.6\%   & 0.1\%  & 0.0\% & 0.0\% \\                 
 \hline             
\end{tabular}
\end{center}

 \begin{center}
\textbf{Table 2. DGP= Geometric (0.1)} \\

\begin{tabular}{lcccccr}
\hline
   n   & & 20 & 60  &   90   &   150  &  250   \\
 \hline
$ \hat{\lambda} $ & & 8.9831 &    8.9340     & 8.9504  & 8.9955 & 8.9791\\
                 & & (2.0556) &   (1.2048)  & (1.0019) &  (0.7643)   & (0.5925)     \\
                 &   &           &          &          &             &          \\
$ \hat{\theta} $  & &  0.10458 &  0.1021   &  0.1015 &  0.1006  & 0.1005   \\
                & & (0.02241) &   (0.0123)&  (0.0102) &   (0.0077)  & (0.0059)           \\
                 &   &           &          &          &             &          \\
$ \hat{  \mathcal{D}}_{KL_{1}} $ & &  0.1163   & 0.4183 & 0.6827 & 1.2375 &2.2480 \\
                     & & (0.0732)&   (0.1359)     & (0.1852) & (0.2660) & (0.4032)  \\
                                          &   &           &          &          &      \\                              
$ \hat{ \mathcal{D}}_{KL_{2}} $ & & 0.5433     & 1.2596 & 1.8269  & 2.9174 &4.9253\\ 
                         & & (0.5215)    & 0.6907) & (0.9123) &  (1.0476) & (1.4205)  \\
                         &   &           &          &          &             &       \\                        
$ KL_{n} $ & & -0.8593 &  -1.2391    & -1.3335  &   -1.7367 & -2.1483 \\
     &  & (2.2220) &  (5.2589)     &  (8.1391)   & (11.8464) & (19.7045)              \\                     
Model selection & Correct       & 11\%    & 19.5\%  & 19.4\% & 34.6\% &  50.4\%\\
based on $ KL_{n} $  & Indecisive  &  89\% & 80.5\%  & 80.6\%  &   65.4\% & 49.6\%\\
                & Incorrect   &  0.0\%  & 0.0\%   & 0.0\%  & 0.0\% & 0.0 \%\\                 
 \hline             
\end{tabular}
\end{center}
\vspace*{0.1cm}

\newpage

\vspace*{0.1cm}

\begin{center}
\textbf{ Table 3. DGP=  0.25 Poisson (9) + 0.75 Geometric (0.1)}  \\

\begin{tabular}{lcccccr}
\hline
   n   & & 20   &   60   &   90  &  150 & 250   \\
 \hline
$ \hat{\lambda} $ & &  8.6053 &    8.5710     & 8.5857 & 8.6201 & 8.6073 \\
                 & & (1.5539) &   (0.9112)  & (0.7599) &  (0.5790) & (0.4481)        \\
                 &   &           &          &          &             &          \\
$ \hat{\theta} $  & &  0.1068     &   0.1054   & 0.1049 &  0.1043 &0.1043    \\
                & & (0.0173)    & (0.0099)&  (0.0082) &   (0.0062) & (0.0048)     \\
                 &   &           &          &          &             &          \\                
$ \hat{ \mathcal{D}}_{KL_{1}} $ & &  0.0944   & 0.3112 &  0.5147 &  0.9447 & 1.7313 \\
                        & & ( 0.0696)  & ( 0.1047) &  (0.1456) &  (0.2234)& ( 0.3607)\\
                        &   &           &          &          &         &          \\                 
$ \hat{ \mathcal{D}}_{KL_{2}} $ & &  0.2619 &   0.6287 & 0.9044 & 1.5169 & 2.5895 \\
                          & & (0.2344)  & (0.2725) &  (0.3477) & ( 0.4544)& ( 0.6205)\\
                          &   &           &          &          &                      \\
$ KL_{n} $ & & -0.8308   & -1.1985 & -1.1990  &   -1.4620 & -1.7481   \\
           &  & (0.9019) & (2.0524) &  (3.0837)   & (4.7931) & (7.7619)   \\                     
Model selection & Geometric       & 9.7\%  &   17\%   & 16.5\%  &  22.4\% &  33.8\% \\
based on $ KL_{n} $  & Indecisive  &  90.3\% & 83\% &  83.5\%  & 77.6\%  & 66.2\%\\
                     & Poisson  &  0.0\% &  0.0\% & 0.0\%   & 0.0\%  & 0.0\% \\                 
 \hline             
\end{tabular}
\end{center}
\vspace*{0.1cm}

\begin{center}
\textbf{Table 4. DGP= 0.5 Poisson (9) + 0.5 Geometric (0.1)}

\vspace*{0.1cm}

\begin{tabular}{lcccccr}
\hline
   n   & & 20  &   60   &   90  &  150 & 250  \\
 \hline
$ \hat{\lambda} $ & & 8.7256   &   8.7074    & 8.7229  &  8.7445 & 8.6073 \\
                 & & (1.0837)   & (0.6383)  & (0.5351) &  (0.4048) & (0.4481)        \\
                 &   &           &          &          &             &          \\
$ \hat{\theta} $  & &  0.1040    &   0.1034   & 0.1031 &  0.1027 & 0.1043    \\
                & & (0.0115)    & (0.0067)&  (0.0056) &   (0.0042) &  (0.0048)            \\
                 &   &           &          &          &             &          \\
$ \hat{  \mathcal{D}}_{KL_{1}} $ & &  0.1488  &   0.4729 & 0.7290 &  1.2969 &  2.3195 \\
                         & &  (0.0665)  & (0.1321) & (0.1806) & (0.2681)& (0.4092)\\
                            &   &           &       &          &             &          \\
$ \hat{ \mathcal{D}}_{KL_{2}} $ & &   0.0873  & 0.1865 &  0.2730 &  0.4921 & 0.8988  \\
                        & &   (0.0716)  & (0.0885) & (0.0902) & ( 0.1435)& (0.2424)\\
                            &   &           &          &          &                \\                            
$ KL_{n} $ & & 0.7491    & 2.3460 & 2.8037 &   3.3866 &  3.9257   \\
          &  & (0.3675)     & (0.9455) &  (1.5428)   & (2.9105) & (5.7219)               \\                     
Model selection &  Geometric       & 1.3\%  & 0.1\%   & 0.0\%  & 0.0\% & 0.0\% \\
based on $ KL_{n} $  & Indecisive  & 92.8\% &  37.1\% & 20.6\%  &  5.3\%  &    1.2\%\\
                & Poisson   &  5.9\% &  62.8\% & 79.4\%   &  94.7\%  & 98.8\% \\                 
 \hline             
\end{tabular}
\end{center}

\newpage

\begin{center}
\textbf{Table 5. DGP=0.75 Poisson (9)+ 0.25 Geometric (0.1)} \\

\begin{tabular}{lcccccr}
\hline
   n   & & 20 &   60   &   90  &  150 & 250  \\
 \hline
$ \hat{\lambda} $ & & 8.6021 &  8.5956     & 8.9967  & 8.6201 &  8.6141 \\
                 & & (0.7146) &  (0.4257)  & (0.3583) &  (0.2656) & (0.2081)       \\
                 &   &           &          &          &             &          \\
$ \hat{\theta} $  & &  0.1047    &    0.1044   & 0.1042 &  0.1040 & 0.1040   \\
                & & (0.0078) &   (0.0046)&  (0.0038) &   (0.0028) &  (0.0022)            \\
                 &   &           &          &          &             &          \\
$ \hat{  \mathcal{D}}_{KL_{1}} $ & &     0.1298 &  0.4627 & 0.7190 &  1.2914 & 2.3190 \\
                         & & (0.0374)  & (0.0954) & (0.1345) & (0.2211) & (0.3751)\\
                            &   &           &          &          &             &    \\
$ \hat{ \mathcal{D}}_{KL_{2}} $ & &  0.1074 &     0.1558 & 0.2060 & 0.3037 & 0.4972 \\
                        & &   ( 0.0917)  & (0.0939) & (0.1117) & (0.1312) & (0.1641)\\
                         &   &           &          &          &                     \\                            
$ KL_{n} $ & & 0.2273   & 2.8437 & 3.7625  &   4.6600 & 5.4338  \\
     &  & (0.4395)     & (0.8358) &  (1.2935)   & (2.5960) &  (5.3012)            \\                     
Model selection & Geometric       & 3.9\%  & 0.2\%   & 0.1\%  & 0.0\% & 0.0\% \\
based on $ KL_{n} $  & Indecisive  & 95.2\% & 14.2\% & 1.5\%  & 0.4\%  &    0.0\%\\
                & Poisson   &  0.9\% &  85.6\% & 98.4\%   & 99.6\%  & 100\% \\                 
 \hline             
\end{tabular}
\end{center}
\vspace*{0.2cm}

The first half of each table gives the average values of the  MLE  $ \hat{\lambda}$ 
and $ \hat{\theta} $, the Kullback-Leibler
divergence test statistics $ \hat{  \mathcal{D}}_{KL_{1}} $ and $ \hat{ \mathcal{D}}_{KL_{2}} $ and the model selection
statistic $ KL_{n} $. The second half of each table gives 
in percentage the number of times our proposed model selection procedure based on $ KL_{n} $ favors the Poisson
model, the Geometric model and indecisive. The tests are conducted at 5\% nominal significance level. 
In the first two sets of experiments ($ \pi= 0.00 $ and $\pi= 1.00$ ) where the model is correctly specified, 
we use the labels \textit{correct, incorrect} and \textit{indecisive} when a choice is made. The first halves 
of \textbf{Tables 1-5} confirm our asymptotic results. They all show that the  MLE  
 $ \hat{\lambda}$ and $ \hat{\theta} $ converge rapidly to their true values in the correctly specified cases
and to their pseudo-true values in the misspecified cases  as the sample size increases. The statistics $ \hat{  \mathcal{D}}_{KL_{1}}$ 
and $ \hat{ \mathcal{D}}_{KL_{2}} $  increase at the rate of $n$, as expected when the models are 
correctly specified and  when the models are  misspecified. As expected, our statistic divergence 
$KL_{n}$ converge to $-\infty$ ( \textbf{Tables 2 } and \textbf{3} ) and to $+\infty $ ( \textbf{Tables 1,4}
 and \textbf{5}) as the sample size increases. \\
 Turning to the second halves of  \textbf{Tables 1} and \textbf{2}, 
 we  note that the percentage of correct choices using model selection statistic steadily increases 
and ultimately converge to  100\%. As a consequence, the probability of correct choice (PCS) based on
Monte Carlo simulation is found to be significantly higher in choosing the correct model in this selection 
procedure based on KLD.  This preceding comments can be applied to the second halves of \textbf{Tables 3}, \textbf{4} and \textbf{5}. 
In all tables, as sample size increases, the percentage of incorrect model  still keeping the same.i.e. 0.0\%.
This is because in KLD, the correct model represents 
the "true" distribution of observations while the incorrect model represents an approximation of the true model.
Except in \textbf{Tables 4} the percentage of incorrect model converges to zero. This is because  the Poisson and
Geometric distributions are approximately at equal distance to the mixture $ t(\pi) $ according
to statistics $ \hat{ \mathcal{D}}_{KL_{1}}$ and $ \hat{ \mathcal{D} }_{KL_{2}}$.\\
 For $n=90$, we plot the histogram of datasets and overlay the curves for Poisson and Geometric distributions. 
 As can be observed in \textbf{Figure 1} and \textbf{3}, the Geometric distribution is distinguished 
from Poisson distribution  and is closely approximates the data sets. In \textbf{Figure 2} and \textbf{5} 
the Poisson distributions is much closer to the data sets. When $\pi=0.5$  ( \textbf{Figure 4} )
the  Poisson distribution and Geometric distribution try to be closer to the data set.
This follows from the fact that they are equidistant from the DGP.

\begin{figure}[htbp]
\centering
\begin{minipage}{.5\textwidth}
  \centering
  \includegraphics[height=4cm,width=\linewidth]{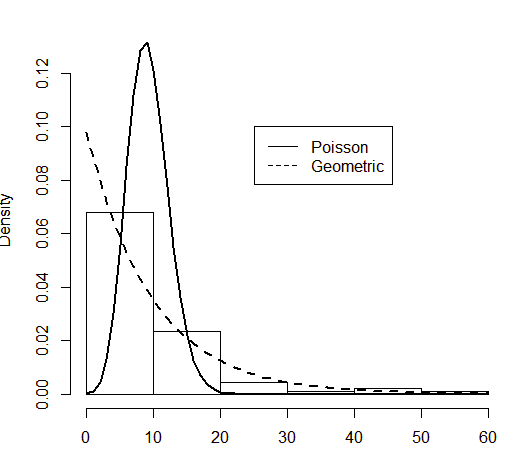}\\
  \caption{{\footnotesize Histogram of DGP = Geom(0.1).  \label{fig1}}}
\end{minipage}%
\begin{minipage}{.5\textwidth}
  \centering
  \includegraphics[height=4cm,width=\linewidth]{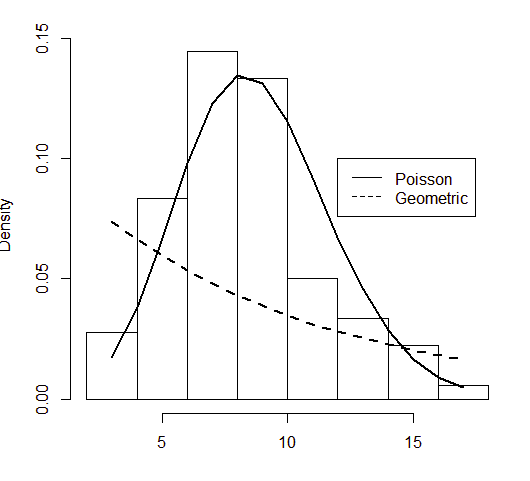}\\
  \caption{{\footnotesize Histogram of DGP = Pois(9). \label{fig2ch4}}}
\end{minipage}
\end{figure}

\begin{figure}[htbp]
\centering
\begin{minipage}{.5\textwidth}
  \centering
  \includegraphics[height=4cm,width=\linewidth]{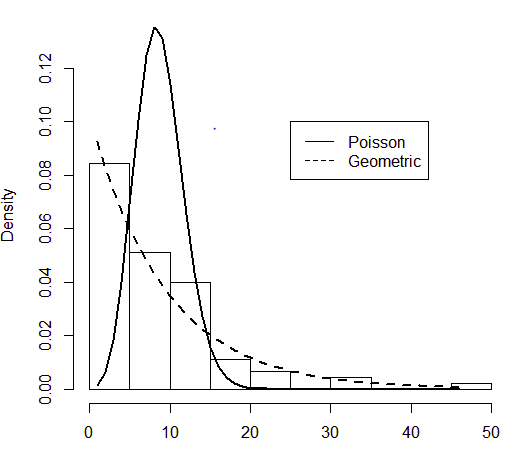}\\
\caption{ {\footnotesize Histogram of DGP = 0.25 Pois(9) + 0.75Geom(0.1). } }
\end{minipage}%
\begin{minipage}{.5\textwidth}
  \centering
  \includegraphics[height=4cm,width=\linewidth]{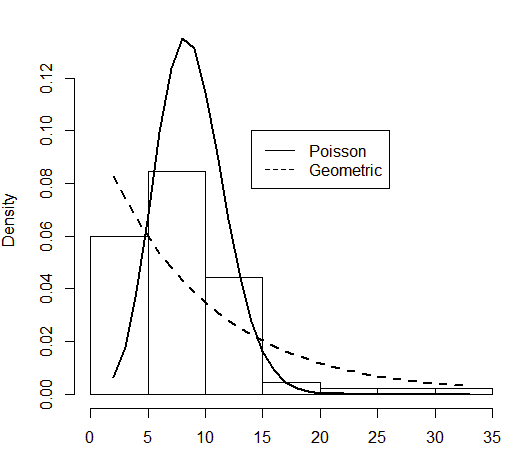}\\
  \caption{{\footnotesize Histogram of DGP = 0.5Pois(9) + 0.5 Geom(0.1). \label{fig2ch4} }}
\end{minipage}
\end{figure}

\begin{figure}[htbp]
\centering
  \includegraphics[height=5cm,width=\linewidth]{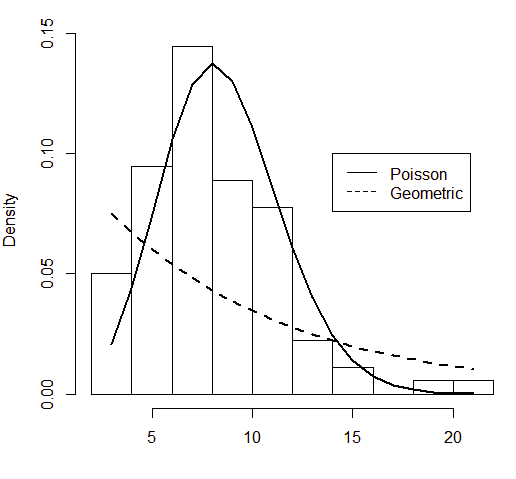}\\
  \caption{{\footnotesize Histogram of DGP = 0.75 Pois(9) + 0.25
Geom(0.1).  \label{fig1}}}
\end{figure}

\newpage

\section{Conclusion}
This paper  shows the strong consistency  of the bias reduced kernel density estimator and establishes a
strongly consistent  Kullback-Leibler divergence estimator based on the bias reduced kernel density estimator.
Furthmore, we have considered an application  in  the problem of selecting estimated models using
Kullback-Leibler divergence estimator  type statistics proposed. 
Specifically, we have proposed some asymptotically standard normal and hypothesis testing  based 
on  Kullback-Leibler divergence estimator constructed in terms of the  bias reduced 
kernel density estimator. We have also proposed  a cross-validation
bandwidth selection for the bias reduced kernel density estimator. 
Our tests are based on testing whether the candidate models are equally close to the true distribution
against the alternative hypotheses that one model is closer than the other where closeness of a model 
is measured according to the discrepancy implicit in the Kullback-Leibler divergence type statistics used. 
The simulations studies show  that the Kullback-Leibler divergence estimator based on the bias reduced
kernel density estimator is  more efficient estimator of  Kullback-Leibler divergence  than 
the Kullback-Leibler divergence estimator based on an ordinary kernel density estimator. 
 The model selection procedure based on the Kullback-Leibler divergence estimator proposed  is competitively especially in small samples.
 Since the Kullback-Leibler divergence is a special case of  $f$-divergences as well as the class of
 Bregman divergences. It would be interesting to propose the methods for discrimations
based on others divergence measures.
\section*{Acknowledgements}
This research was supported, in part, by grants from CEA-SMA Centre d'Excellence Africain en Sciences\\
Math\'ematiques et Applications IMSP, B.P. 613 Porto-Novo, Benin. The authors are grateful for helpful comments from Carlos OGOUYANDJOU.

\end{document}